\documentclass[12pt]{article}
\usepackage{epsfig,amsfonts,amssymb}
\usepackage{amsmath,graphics}

\usepackage{hyperref}
\def\ZZZ{{\hbox{ Z\kern-1.6mm Z}}}
\def\RRR{{\hbox{ R\kern-2.4mm R}}}
\def\CCC{{\hbox{ C\kern-2.0mm C}}}
\def\zzz{{\hbox{z\kern-1mm z}}}
\newcommand{\mathsym}[1]{{}}

\newcommand{\qeq}{{\hbox{=\kern-2.3mm ? \kern.5mm }}}
\renewcommand{\qeq}{=}

\newcommand{\be}{\begin{equation}}
\newcommand{\ee}{\end{equation}}
\newcommand{\ben}{\begin{eqnarray}\displaystyle}
\newcommand{\een}{\end{eqnarray}}

\def\one{{\hbox{ 1\kern-.8mm l}}}
\def\zero{{\hbox{ 0\kern-1.5mm 0}}}

\begin{document}

\baselineskip 24pt

\begin{center}
{\Large \bf On the smoothness of  multi center coplanar black hole and membrane horizons.}

\end{center}

\vskip .6cm
\medskip

\vspace*{4.0ex}

\baselineskip=18pt

\centerline{\large \rm   Chethan N. Gowdigere, Abhass Kumar \footnote{Present address: Harish-Chandra Research Institute, Chhatnag Road, Jhunsi, Allahabad 211019;  abhasskumar@hri.res.in }, Himanshu Raj \footnote{Present address: International School for Advanced Studies,
via Bonomea 265, 34136 Trieste, Italy; hraj@sissa.it} and Yogesh K. Srivastava}

\vspace*{4.0ex}

\centerline{\large \it National Institute of Science Education and Research.}

\centerline{\large \it  Sachivalaya Marg, PO: Sainik School,}

\centerline{\large \it  Bhubaneswar 751005, INDIA}

\vspace*{1.0ex}
\centerline{E-mail: chethan.gowdigere,  yogeshs@niser.ac.in }

\vspace*{5.0ex}

\centerline{\bf Abstract} \bigskip
We study the differentiability of the metric and other fields at any of the horizons  of  multi center Reissner-Nordstrom black hole solutions in $d \ge 5$ and of multi center $M2$ brane solutions.  The centers are distributed in a plane in transverse space, hence termed coplanar. We construct the Gaussian null co-ordinate system for the neighborhood of a  horizon by solving the geodesic equations in expansions of (appropriate powers of) the affine parameter. Organizing the harmonic functions that appear in the solution in terms of what can be called generalized Gegenbauer polynomials is key to obtaining the solution to the geodesic equations in a compact and manageable form.  We then compute the metric and other fields in the Gaussian null co-ordinate system and find that the differentiability of the coplanar solution is \emph{identical to} the differentiability of the collinear solution (centers distributed on a line in transverse space).  The results of this paper thus run counter to a suggestion in the literature that posits  reduction in the degree of smoothness to accompany  reduction in symmetries. We end the paper with a  conjecture on the degree of smoothness of the most general multi center solution, the one with centers distributed arbitrarily and hence  possessing no transverse spatial isometries. 

\vfill \eject

\baselineskip=18pt

\tableofcontents

\section{\label{1}Introduction and Motivation} \label{s1}
In this paper, we study horizon smoothness of multi-center black hole solutions in various dimensions and of multi-membrane solutions in M-theory. By studying horizon smoothness is meant the determining of the degree of differentiability (smooth being $\mathcal{C}^\infty$, only $k$-times differentiable $\mathcal{C}^k$) at the horizon of the (components of the) various fields present in the solution such as the metric, gauge fields, tensor gauge fields.  Typically,\footnote{Although, more recently, black hole solutions have been discovered differently \cite{Gutowski:2004ez}: one first solves the equations and determines the various possible near horizon solutions, subsequently then one solves the equations to obtain a solution that interpolates between a near horizon solution and the asymptotic solution.} a black hole solution is obtained by solving the Einstein's equations in a co-ordinate patch that includes asymptotic infinity and is adapted to describe observers located outside the horizon. The (components of the) fields of the solution are smooth functions in that co-ordinate patch. Such a co-ordinate patch,usually, does not cover the horizon and space-time regions interior to it. An acceptable black hole solution in classical gravity should have all fields sufficiently differentiable(for example, one usually demands that metric be twice differentiable ) everywhere i.e. at  all points including and especially at the horizons, except, of course, at the curvature singularities located within the horizons. From the first presentation of the solution i.e.  in the co-ordinate patch adapted to outside the horizon observers, it is not clear if one has an acceptable black hole solution or not, in particular, if the fields of the solution have required degree of differentiability at the horizon, which is the main concern of this paper.  To address  such questions, one has to extend the solution by finding another set of coordinates which are well-defined at the horizon and in it's neighborhood.

 In this paper, following \cite{Candlish1}, we work in a Gaussian null co-ordinate system.The Gaussian null co-ordinate system is constructed using null geodesics: the solution of the geodesic equations (written out in the co-ordinate system of the first presentation of the solution) provides the smooth transition functions between the co-ordinate patch of the first presentation and a co-ordinate patch that covers the  horizon and it's neighborhood, henceforth referred to as the Gaussian null co-ordinate patch.  The transition functions between the two co-ordinate patches are then used to obtain the component functions of the various tensor fields of the solution in the Gaussian null co-ordinate patch via the tensor transformation law. From the expressions of the component functions in the Gaussian null co-ordinate patch, we then examine for their smoothness or lack of it at the horizon and in the latter case read off the degree of smoothness; we do this for all components of a tensor field and the tensor field is said to be $\mathcal{C}^m$ when each of the components is at least $m$ times differentiable (there could very well be some components which are differentiable more than $m$ times). We say that a horizon is smooth if all components of all tensor fields of the solution are smooth at the horizon; else we say that the horizon is not smooth and further supplement the statement by giving the degree of smoothness of the various tensor fields of the solution.

In this article, we examine  two classes of solutions for horizon smoothness. The first class is of the electrically  charged black hole solutions of Einstein-Maxwell theory in $d\geq5$ space-time dimensions. Equation \eqref{emaction} contains the action of Einstein-Maxwell theory in $d$ dimensions  and equations \eqref{rnsoln}, \eqref{rnsoln2} contain the (first presentation of the) black hole solutions we will study (the co-ordinate system in the first presentation is known as the isotropic co-ordinate system); note that these formulae are valid for $d \geq 4$. The second class of solutions is that of electrically charged membrane solutions in M-theory;  equations \eqref{m2soln}, \eqref{m2harm} give the (first presentation of the)  solution in isotropic co-ordinates. One common feature of both classes of solutions is that a part of the space-time is conformally a Euclidean space; conformal $\mathbf{R}^{d-1}$ in the black hole case and a conformal $\mathbf{R}^{8}$ in the membrane case, often referred to as the transverse Euclidean space. Furthermore each of these solutions is completely specified by a harmonic function, the $H$ that appears in the equations \eqref{rnsoln2} and \eqref{m2harm}, harmonic in the transverse Euclidean space. When $H = 1 + \frac{\mu}{r^{d-3}}$ in the black hole case and when $H = 1 + \frac{\mu}{r^6}$ in the membrane case, the solutions describe a single black hole and a single stack of membranes respectively; these solutions are referred to as  single center solution because the harmonic function has a term due to a single center (of charge). The single center is located at the origin of the transverse Euclidean space; the origin itself is not part of the isotropic co-ordinate patch for the single center solution and as one approaches the origin one in fact approaches the horizon. One (relevant) part of the isometries of the solution are generated by Killing vector fields corresponding to rotations around the origin in the transverse Euclidean case. In the black hole case these spatial rotational symmetries constitute a  $\mathbf{so}(d-1)$ while in the membrane case they constitute a $\mathbf{so}(8)$. The only horizon of the single center solutions is known to be smooth; we will be able to see this as a special  case of the results of this paper.

By adding more centers to the harmonic function, one obtains solutions describing multiple black holes or multiple membranes; any two of them are in static equilibrium due to the cancellation of gravitational and electric forces which happens because of the equality (in appropriate units) of their masses and charge. We thus have solutions with multiple horizons. When we have two centers, one finds that the spatial rotational symmetries of the solution are only those rotations in the transverse Euclidean space that preserve the line joining the two centers. This is true even for an arbitrary number of centers all located on one line. The spatial rotational symmetries constitute a $\mathbf{so}(d-2)$ in the black hole case and a $\mathbf{so}(7)$ in the membrane case. We will refer to this situation sometimes as  the ``two center'' case, even if in actuality there are an arbitrary number of centers, because two is the smallest number that has this symmetry, and also as the ``collinear'' case. Going on, when we have three centers or even an arbitrary number (greater than three) of centers all distributed on one plane in the transverse Euclidean space, the spatial rotational symmetries of the solution constitute a $\mathbf{so}(d-3)$ in the black hole case and a $\mathbf{so}(6)$ in the membrane case. We will refer to this situation as the ``three center'' case or alternatively as the ``coplanar'' case. All the computations in this paper concern the three center/coplanar situation. Going further on, the $k$-center case corresponds to a spatial rotational symmetry $\mathbf{so}(d-k)$ in the black hole case and a $\mathbf{so}(9-k)$ in the membrane case; of course $k \leq d$ in the black hole case and $k \leq 9$ in the membrane case because for larger number of centers in arbitrary positions all spatial rotational symmetries are broken. The discussion in the present paragraph on symmetries of multi center solutions is pertinent to the discussion on the degree of smoothness of these solutions to be made subsequently.

Now that one has obtained solutions that describe multiple black holes and multiple membranes, solutions with multiple horizons, one needs to investigate if these solutions are acceptable solutions, in particular if the horizons, each one of them,  are smooth. A horizon when it occurs by itself in the single center solution is smooth, does it continue to be smooth when it occurs in the presence of other horizons, as it does in the multi center solutions?  The first investigation of this kind was carried out for the the Majumdar-Papapetrou solutions, which are nothing but the $d=4$ multi center black hole solutions in equation \eqref{rnsoln}.  Hartle and Hawking were able to obtain a horizon co-ordinate system in a somewhat direct manner:  a co-ordinate transformation from the isotropic co-ordinates gives new co-ordinates which upon extending the ranges of the co-ordinates covers the first horizon\footnote{$H = 1 + \frac{\mu_1}{r} + \frac{\mu_2}{\| \vec{r} - \vec{r}_2\|} + \frac{\mu_3}{\| \vec{r} - \vec{r}_3\|} +\ldots$; the first horizon is at $r=0$.}. On computing the metric and gauge fields in this co-ordinate system, they could ascertain that the (first) horizon is smooth. This of course means that each of the multiple horizons is similarly smooth. The analysis of the smoothness of horizons for the $d \geq 5$ black hole case was done in \cite{Candlish1} by Candlish and Reall, building on earlier work by \cite{Welch}. The membrane solutions were analyzed by some of us in \cite{Gowdigere:2012kq}; the first indication that they may not be smooth was there in \cite{Gibbons}. In \cite{Candlish1}, they  considered only the collinear case and the horizon co-ordinate system they worked with is the Gaussian null co-ordinate system, whose construction we have already briefly reviewed above. We will give a brief account of the results of \cite{Candlish1} because (one part of) the present work is built on and extends it.

Introducing co-ordinates on the transverse Euclidean space in the following manner, 
\ben 
x_1 &=& r  \cos \theta, \nonumber \\
x_2 &=& r \sin \theta \, \cos \phi, \nonumber \\
x_3 &=& r \sin \theta \, \sin \phi \, \cos \psi_1, \nonumber \\
x_4 &=& r \sin \theta \, \sin \phi \, \sin \psi_1 \, \cos \psi_2, \nonumber \\
\vdots \nonumber \\ 
x_{d-2} &=& r \sin \theta \, \sin \phi \, \sin \psi_1 \, \sin \psi_2  \,\ldots  \sin \psi_{d-5}\, \cos \psi_{d-4}, \nonumber \\ 
x_{d-1} &=& r \sin \theta \, \sin \phi \, \sin \psi_1 \, \sin \psi_2  \,\ldots  \sin \psi_{d-5}\, \sin \psi_{d-4}, \label{spcoordinates}
\een
the harmonic function for the collinear black hole solution is
\be \label{1.2} H(r, \theta) = 1 + \frac{\mu_1}{r^{d-3}} + \sum_{i=2}^\infty \frac{\mu_i}{(r^2 - 2 \,r\, b_i  \cos \theta+ b_i^2)^{\frac{d-3}{2}}}. \ee 
All the black hole centers are located on the $x_1$-axis passing through the origin; the first black hole (with charge $\mu_1$) is located at the origin and the others (with charges $\mu_i$) are on this axis at $x_1 = b_i$.  We will see later that for the analysis of the geodesic equations, due to the boundary conditions imposed, it is convenient to organize the harmonic function as a series expansion in $r$: 
\be  \label{1.3} H(r, \theta) = \frac{\mu_1}{r^{d-3}} +  \sum_{n=0}^\infty r^n \, h_n \, G_n(\cos \theta).\ee 
In the above formula, $G_n(y)$ are Gegenbauer polynomials, defined by their generating function 
\be \label{gpgenerating} \frac{1}{(1- 2yz + z^2)^{\frac{d-3}{2}}} =  \sum_{n=0}^\infty  z^n \,G_n(y) \ee
and the $h_n$'s are constants that depend on the parameters of the black holes viz. the charges $\mu_i$ and their locations $b_i$
\be\label{hn} h_n = \delta_{n,0} + \sum_{i=2}^\infty \frac{\mu_i}{~b_i^{n+d-3}}.\ee
Note that in the harmonic function \ref{1.3}, the parameters of the black holes other than the first one are contained only in the $h$'s and nowhere else; hence the difference between two collinear black holes and any number of collinear black holes lies only in the $h$'s and nothing else.

The construction of the Gaussian null co-ordinate system for the neighborhood of a  horizon (first black hole's) requires the solution to the radial null geodesic equations.  Due to the symmetries of the collinear solution, only three of the co-ordinates are non-trivial along the geodesic viz. $t(\lambda), r(\lambda)$ and $\theta(\lambda)$ with $\lambda$ being the affine parameter along the geodesic; the solution for the other co-ordinates is simply $\phi(\lambda)=$ constant, $\psi_1(\lambda)=$ constant, $\ldots$ $\psi_{d-4}(\lambda)=$ constant. One needs to solve only the geodesic equations corresponding  to the $r$ and $\theta$ co-ordinates (i.e. those equations in which the second derivatives of $r(\lambda)$ and $\theta(\lambda)$ occur),  the solution to $t(\lambda)$ then follows from the  $\frac{\partial}{\partial t}$  Killing symmetry of the metric; hence to construct the Gaussian null co-ordinate system for the collinear case, one only has to deal with two coupled non-linear o.d.e's   for two functions instead of the apriori $d$ coupled non-linear equations for $d$ functions.  The co-ordinates of the Gaussian null co-ordinate system are $\lambda, v, \Theta, \Phi, \Psi_1, \ldots \Psi_{d-4}$; $\lambda$ is the affine parameter, $v$, $\Theta$, $\Phi$ and the $\Psi$'s are constants of integration that appear in the solution to the geodesic equations, see  \cite{Candlish1} or the sequel for details. The solution to the geodesic equations are then thought of as  the transition functions from the isotropic co-ordinate system to the Gaussian null co-ordinate system. Candlish and Reall \cite{Candlish1} then compute the components of the metric and the gauge field in the Gaussian null co-ordinate system to find that they are not smooth at the horizon. For $d=5$, the metric was found to be only $\mathcal{C}^2$ and the gauge field was found to be $\mathcal{C}^0$ and for $d \ge 6$ the metric was found to be only $\mathcal{C}^1$ and the gauge field $\mathcal{C}^0$.

The analogous analysis of  the degree of  smoothness of the membrane horizon for the collinear case was done in \cite{Gowdigere:2012kq} by some of us. Constructing the Gaussian null co-ordinate system is a little more involved owing to the peculiarities of the membrane horizon (in comparison to a hole horizon); the details of this construction will be reviewed later in this work. It was shown in \cite{Gowdigere:2012kq} that the metric is only $\mathcal{C}^3$  while the tensor gauge field strength is only $\mathcal{C}^2$ at the horizon. 

Having finished with describing the status of previous work on the analysis of smoothness of horizons, we can now state what we intend to study in this paper. We will study thedegree of smoothness of horizons of  three-center/coplanar configurations, both for  black holes and membranes. One motivation for such a study is simply to take the next step towards completing the task of studying arbitrary multi-centre configurations; after all the two center/ collinear case is not the most general configuration, it being a starting point due to its simplicity.  One wishes to answer questions such as, is the degree of smoothness of  such configurations less than or equal to that of the collinear configuration. Another motivation comes from the work of Candlish \cite{Candlish2}, where a certain conclusion is drawn from various studies on the smoothness of  multi horizon solutions. 

\subsubsection {\label{101}A connection between differentiability and symmetry}

Following is an exact line from \cite{Candlish2}:  ``The lack of smoothness present  for higher dimensional black holes seems to be ubiquitous in situations where rotational symmetries of the single black hole solution are broken.''  A connection is alluded to between the fact that the multi horizon solutions break some of the spatial isometries/ rotational symmetries of the single horizon solution and the fact that the degree of smoothness of the  horizons of the multi horizon solutions is less than the degree of  smoothness of the horizon of the single horizon solution. The first example to offer towards this connection concerns the collinear black holes and membranes that we have discussed so far: the multi horizon configurations break the single centre symmetry from $\mathbf{so}(d-1)$ to $\mathbf{so}(d-2)$ and from $\mathbf{so}(8)$ to $\mathbf{so}(7)$ respectively and there is an accompanying reduction in smoothness. The second example to offer in favor of the connection concerns the concentric black ring solutions of  \cite{Gauntlett:2004wh}.  It turns out that, partly because all the rings are concentric and in the same plane,  the multi ring solution preserves all the isometries of the single ring solution;  and consequently there should not be any loss of smoothness and indeed it was shown in \cite{Gauntlett:2004wh} that each of the ring horizons is smooth. All further examples  involve the BMPV black hole, which is a five dimensional rotating black hole; it rotates in both planes and it preserves at least a  $U(1)_\rho \times U(1)_\psi$ symmetry, where $\rho$ and $\psi$ are the angular co-ordinates in the two planes. The third example that can be given towards the connection is the multi horizon solution constructed also in \cite{Gauntlett:2004wh}; this solution has  a BMPV black hole at the centre of the concentric rings in the plane of the rings.  This example can be thought of in two different ways, both favoring the connection alluded to by Candlish. The first is to think of the single horizon solution to be the BMPV black hole and the multi horizon solution to be the hole + rings solution. Since the hole + rings solution preserves all of the $U(1)_\rho \times U(1)_\psi$ isometries of the BMPV black hole solution, the horizon of the BMPV black hole should be smooth and indeed it is as was shown in \cite{Gauntlett:2004wh}. The second way to think of the hole + rings solution is to think of the single horizon solution to be one of the black rings. Again the hole + rings solution preserves all the isometries of the single black ring solution and hence  the horizon of the black ring should be smooth and indeed that is the case.  The fourth example that seems to strengthen the connection between reduced symmetries and reduced smoothness concerns the hole + ring solutions of \cite{Bena:2005zy}. The multi horizon solutions presented in \cite{Bena:2005zy} is that of a BMPV black hole  and a black ring;  the BMPV black hole is located not at the centre of the black ring in the plane of the ring as in the previous example but displaced in the plane perpendicular to the plane of the black ring at the centre of the black ring. The symmetries of this multi-horizon solution is less than that of the individual BMPV black hole as well as as that of the individual black ring solution. From the connection alluded to by Candlish, one should expect that the black hole horizon as well as the black ring horizon in the multi horizon solution should not be smooth. Candlish in \cite{Candlish2} argues that this is indeed the case. The fifth example that favors the connection concerns the multi black hole solutions studied by Candlish in \cite{Candlish2}, which involve a BMPV black hole together with a line of static black holes. The multi horizon solution breaks one of the two $U(1)$ isometries of the single horizon BMPV solution and consequently there should be a loss in the degree of smoothness of the BMPV horizon; Candlish computes this and finds that the metric is only ${\cal C}^2$ and the gauge field is only ${\cal C}^0$.

The connection between differentiability and smoothness that has been reviewed above raises some pertinent questions for the three center/ coplanar configurations. It is clear that the horizons are not expected to be smooth  since they break the isometries of the single horizon solution (also simply because the collinear solution is a special case of the coplanar solution.)  But the question is the smoothness of the coplanar horizons in comparison to the collinear horizons. The coplanar solutions break more of the isometries of the single center solution than the collinear solution. Does this mean that the coplanar horizons are less smooth than the collinear horizons?  Is the decrease in the symmetry of the solution accompanied by a decrease in the degree of smoothness of the horizons?  But this cannot be true because if there were a loss of smoothness associated to every step in the decrease of the isometry which happens when we increase the number of centers (in generic positions), we would soon have no differentiability. Perhaps, the fact that the coplanar configuration breaking more isometries than the collinear solution does not matter and  the only thing that matters is that they both break  some isometries of the single center solution. If that is the case, then perhaps the horizons in the coplanar and collinear solutions have the same degree of smoothness; this would then mean that most general multi-center configuration would also have the same degree of smoothness.  The only way to decide which of these speculations is correct is by actually doing the computations. 

Equipped with all the aforementioned motivating questions, we proceed to the computations. The harmonic function for the coplanar/ three centre black hole solution is given by

\be \label{3cenharm}
H(r,\theta, \phi) = 1 + \frac{\mu_1}{r^{d-3}} + \sum_{i=2}^{\infty} \frac{\mu_i}{(\, r^2 - 2\, r\, b_i \cos \theta  - 2\, r\, a_i \sin \theta \cos \phi + b_i^2+a_i^2\,)^{\frac{d-3}{2}}}.
\ee
The black holes are distributed in the $x_1 - x_2$ plane. The first black hole, the one with charge $\mu_1$ is at the origin in the transverse Euclidean space and for whose horizon we construct the Gaussian null co-ordinate system. The other black holes' centers  are located at $(b_i, a_i, 0, \ldots , 0), \quad i = 2, 3, \ldots$ in the transverse Euclidean space. 
Due to the additional explicit dependence on the angle $\phi$ and the accompanying reduction in symmetry, the geodesic equations are more involved compared to the collinear case. We now have four functions $t(\lambda)$, $r(\lambda)$, $\theta(\lambda)$ and $\phi(\lambda)$ non-trivial along the geodesic. Again the solution for $t(\lambda)$ can be found in terms of the others due to the $\frac{\partial}{\partial t}$ Killing symmetry. We need to solve the geodesic equations corresponding to the $r$, $\theta$ and $\phi$ co-ordinates; three coupled non-linear o.d.e's for three functions. One can plug in the expression for $H$ given above \eqref{3cenharm} into the geodesic equations and with the aid of computer algebra packages following the procedure laid out in \cite{Candlish1} which we will describe later here, obtain the solution. But the resulting expressions are huge and cumbersome. We found that  if organized in terms of what we call generalized Gegenbauer polynomials, to be defined below, the expressions substantially reduce in size, reaching lengths similar to the collinear case computations. One is led to  surmise that the answer organizes itself in terms of these generalized Gegenbauer polynomials probably because the question or rather the starting point of the computations is itself also organized in terms of them. We are thus led to reorganize the harmonic function \eqref{3cenharm} as in the following paragraph.  

First, define for each black hole other than the first one, 
\ben \label{fR}
f_i(\theta, \phi)  = \frac{b_i}{R_i} \cos \theta + \frac{a_i}{R_i} \sin \theta \, \cos \phi, \qquad R_i = +\sqrt{b_i^2+a_i^2}.
\een
$f_i(\theta, \phi)$ is  the cosine of the angle between the position vector of the $i$'th black hole and $\vec{r}$, the argument of the harmonic function. The harmonic function \eqref{3cenharm} then becomes
\be \label{3cenharm2}
H(r,\theta, \phi) = 1 + \frac{\mu_1}{r^{d-3}} + \sum_{i=2}^{\infty} \frac{\mu_i}{(\, r^2 - 2\, r\, R_i f_i(\theta, \phi) + R_i^2\,)^{\frac{d-3}{2}}}\,.
\ee
Using \eqref{gpgenerating}, we can write this as follows:
\be \label{3cenharm7}
H(r,\theta, \phi) = 1 + \frac{\mu_1}{r^{d-3}} + \sum_{i=2}^{\infty} \sum_{n=0}^{\infty} r^n\,\frac{\mu_i}{R_i^{n+d-3}}\,G_n(f_i(\theta, \phi)).
\ee
Now, we define generalized Gegenbauer polynomials
\be \label{ggpdefn}
{\cal G}_n (\theta, \phi) =  \delta_{n,0} + \sum_{i=2}^{\infty} \frac{\mu_i}{R_i^{n+d-3}}\,G_n(f_i(\theta, \phi)).
\ee
The term generalized Gegenbauer polynomials is meant to indicate the above functions of two variables; it is just a name we employ in this paper. It is not meant to indicate a new special function or anything else; in fact the main ingredient that goes into the construction of the generalized Gegenbauer polynomials is the Gegenbauer polynomial.  We can now  write the $r$-series expansion of the harmonic function \eqref{3cenharm}, \eqref{3cenharm2} as follows:
\be \label{3cenharm3}
H(r,\theta, \phi) = \frac{\mu_1}{r^{d-3}} + \sum_{n = 0}^\infty r^n \, {\cal G}_n\,(\theta, \phi)\,.
\ee

When we go to the collinear limit, i.e. set the $a_i$'s to zero, the $R_i$'s become $b_i$'s, all the $f_i$'s  reduce  to $\cos \theta$  and the generalized Gegenbauer polynomial ${\cal G}_n(\theta, \phi)$  is now only a function of $\theta$ and furthermore (for $n\neq 0$) factors into two pieces one of which is a constant  that contains all the black hole parameters,  $h_n$, and the other the ordinary Gegenbauer polynomial, $G_n(\cos \theta)$ which implies that   \eqref{3cenharm3} reduces to \eqref{1.3}.
Note that the $r$-series expansions of the harmonic function in the collinear \eqref{1.3} and the coplanar \eqref{3cenharm3} are very similar; this similarity will form the basis of a conjecture we will make by the end of the paper about the degree of smoothness of the most general multi centre solutions.  

With \eqref{3cenharm3}, the question  or the starting point is posed in terms of generalized Gegenbauer polynomials of the isotropic angles $\theta$ and $\phi$ and their partial derivatives. The answer, perhaps not surprisingly,  will turn out to be expressed in terms of the the generalized Gegenbauer polynomials of the corresponding Gaussian null co-ordinate angles $\Theta$ and $\Phi$ and their derivatives. The use of these generalized Gegenbauer polynomials is the essential ingredient that makes the computations to determine the degree of smoothness of horizons of three center / coplanar solutions manageable: manageable both in terms of time taken to perform the computations and also in terms of the brevity of the final expressions. 

The rest of this paper is organized as follows. In section \ref{2}, we study coplanar / three center black holes first for $d = 5$ in \ref{21},  and then for all $d \geq 6$ in \ref{22}. We solve the geodesic equations in \ref{211} and \ref{221};  then we obtain the transition functions from the isotropic co-ordinates to the Gaussian null co-ordinate system in \ref{212} and \ref{222}. Subsequently, in \ref{214} and \ref{223}, we compute the components of the various tensor fields in the Gaussian null co-ordinate system and read off the degree of smoothness of the horizon. We then discuss the results in \ref{215} and \ref{224} and obtain  answers to the various motivating questions. In section \ref{3}, we study coplanar / three center $M2$ brane horizons along the same lines as the black hole case and obtain answers to the motivating questions in \ref{304}. In the final concluding section \ref{4}, we first summarize all the results and then try to gather lessons from them for the smoothness of more generic $k$-centre with $ k \geq 4$ solutions. We are able to make a conjecture about the degree of smoothness of the horizons in the most general multi center solution. We also comment about the connection between the loss of symmetries and the loss of smoothness that partly motivated this work.  We are able to offer a different explanation for the lack of smoothness of horizons in multi center solutions which is partly conjecture. We end the paper with directions for future work. We collect some of the longer formulae in appendices \ref{a} - \ref{d}.

\section{\label{2}Three center /  Coplanar Black holes}

The multi center black holes we investigate in this paper are solutions to $d$ dimensional Einstein-Maxwell theory, whose action is given by
\be \label{emaction}
S = \int d^dx~\sqrt{-g}\,\left(R - \frac{d-2}{8(d-3)}\,F_{\mu\nu}\,F^{\mu\nu} \right) \, .
\ee
We are following the conventions of  \cite{Candlish1} here. Following is the first presentation of the black hole solution in  isotropic co-ordinates. The metric and gauge fields are given by
\be \label{rnsoln}
ds^2 = -  H^{-2}\,dt^2 + H^{\frac{2}{d-3}}\, ds^2_{\mathbf{R}^{d-1}}, \qquad  A = -\frac{dt}{H}\, , \ee
where $ds^2_{\mathbf{R}^{d-1}}$ is the flat metric of the transverse Euclidean space $\mathbf{R}^{d-1}$. $H$ is a harmonic function in the transverse Euclidean space:
\be  \label{rnsoln2}H(\vec{r}) = 1 + \sum_{i = 1}^\infty \frac{\mu_i}{\| \vec{r} - \vec{r}_i \|^{d-3}} \, .\ee 
$\vec{r}_i$ are points in the transverse Euclidean space which  correspond to the locations of the horizons of the various black holes and $\| \|$ is the Euclidean norm. We will need to introduce a co-ordinate system for the transverse Euclidean space,  already given in \eqref{spcoordinates}, in which the flat metric  takes the form
\begin{multline} \label{flatmetric}ds^2_{\mathbf{R}^{d-1}} = dr^2 + r^2\,d\theta^2 + r^2\,\sin^2\theta\,d\phi^2 + r^2\,\sin^2\theta\,\sin^2\phi\,d\psi_1^2 +  r^2\,\sin^2\theta\,\sin^2\phi\, \sin^2\psi_1\,d\psi_2^2  + \dots \\ \dots \ldots +  r^2\,\sin^2\theta\,\sin^2\phi\, \sin^2\psi_1 \dots \sin^2\psi_{d-5}\,d\psi_{d-4}^2  \, . \end{multline}
Thus, the co-ordinates in the isotropic co-ordinate system are $t, r, \theta, \phi, \psi_1, \dots \psi_{d-4}$.

In the following, we will first study the five dimensional black holes which behave differently to the six and higher dimensional black holes whose study we take up  subsequently. 

\subsection{\label{21}$d = 5$}
We start by setting $d= 5$ in all formulae appearing in the previous section. In particular, the harmonic function for the coplanar configuration is
\be \label{3cenharm5d}
H(r,\theta, \phi) = \frac{\mu_1}{r^{2}} + \sum_{n = 0}^\infty r^n \, {\cal G}_n\,(\theta, \phi)\,.
\ee
We will not indicate the dimension in the notation for the generalized Gegenbauer polynomials to avoid cluttering; the dimension should be obvious from the context.  The generalized Gegenbauer polynomials appearing in \eqref{3cenharm5d} are the ones built with five dimensional Gegenbauer polynomials.  

To construct the Gaussian null co-ordinate system for the horizon (of the first black hole), we will need to solve for radial null geodesics falling into this horizon.

\subsubsection{\label{211}Solving the geodesic equations}

Due to the $\frac{\partial}{\partial \psi_1}$-Killing symmetry of the metric, the ``$\psi_1$-geodesic'' equation admits a first integral and hence can be readily solved,
\be \label{5dpsigeod} \frac{d}{d\lambda} \left[ H \,r^2\,\sin^2 \theta\,\sin^2 \phi\, \frac{d\psi_1}{d\lambda} \right] = 0 \quad \Longrightarrow \	\quad \psi_1(\lambda) = \Psi_1,\ee
where $\Psi_1$ is an integration constant.  Due to the $\frac{\partial}{\partial t}$-Killing symmetry, the ``$t$-geodesic'' equation admits a first integral which can be solved, 
\be \label{5dtgeod}\frac{d}{d\lambda} \left[ H^{-2} \, \frac{dt}{d\lambda} \right] = 0 ~\Longrightarrow ~ \frac{d}{d\lambda}t(\lambda) = -  H(r(\lambda), \theta(\lambda), \phi(\lambda))^2 ~ \Longrightarrow ~  t(\lambda) = v \,- \int d\lambda \,H(r(\lambda), \theta(\lambda), \phi(\lambda))^2, \ee
where in choosing the integration constant of the first integration  to be $-1$ we have  employed some of the freedom in choosing the affine parameter and $v$ is the second integration constant. Thus, $t(\lambda)$ is determined via \eqref{5dtgeod} in terms of $r(\lambda), \theta(\lambda)$ and $\phi(\lambda)$,  which are obtained by solving simultaneously the ``$r$-geodesic'' equation
\be \label{5drgeod} \ddot{r} - \partial_r H + \frac{\partial_r H}{2 H}\,\dot{r}^2 - \frac{\partial_r H}{2 H}\,r^2\dot{\theta}^2 - r\dot{\theta}^2 - \frac{\partial_r H}{2 H}\,r^2 \sin^2\theta\, \dot{\phi}^2 - r \sin^2\theta\, \dot{\phi}^2  + \frac{\partial_\theta H}{H} \, \dot{r}\,\dot{\theta} + \frac{\partial_\phi H}{H}\,\dot{r}\,\dot{\phi} = 0,\ee
the ``$\theta$-geodesic'' equation
\be \label{5dthetageod}\ddot{\theta} - \frac{\partial_\theta H}{r^2} - \frac{\partial_\theta H}{2 H r^2}\,\dot{r}^2 + \frac{\partial_\theta H}{2 H}\,\dot{\theta}^2  -  \frac{\partial_\theta H}{2 H}\,\sin^2\theta\,\dot{\phi}^2 - \sin\theta \cos \theta\,\dot{\phi}^2 + \frac{2}{r}\,\dot{r} \,\dot{\theta} + \frac{\partial_r H}{H}\,\dot{r}\,\dot{\theta} + \frac{\partial_\phi H}{H}\,\dot{\theta}\,\dot{\phi} = 0\ee
and the ``$\phi$-geodesic'' equation
\be \label{5dphigeod}\ddot{\phi} - \frac{\partial_\phi H}{r^2 \sin^2\theta} - \frac{\partial_\phi H}{2 H r^2 \sin^2\theta}\,\dot{r}^2  - \frac{\partial_\phi H}{2 H  \sin^2\theta}\,\dot{\theta}^2 + \frac{\partial_\phi H}{2 H}\,\dot{\phi}^2 + \frac{2}{r}\,\dot{r} \,\dot{\phi} + \frac{\partial_r H}{H}\,\dot{r}\,\dot{\phi} + 2 \cot \theta \,\dot{\theta}\,\dot{\phi} + \frac{\partial_\theta H}{H}\,\dot{\theta}\,\dot{\phi} = 0.\ee
It is convenient to also consider the null condition which is a consequence of another first integral of the geodesic equations, 
\be \label{5dnull}- H^{-2}\,\dot{t}^2 + H\,\dot{r}^2 + H r^2\,\dot{\theta}^2 + Hr^2\,\sin^2\theta\,\dot{\phi}^2 = 0.\ee

The boundary conditions are chosen as follows. First we employ the remaining freedom allowed in choosing the affine parameter so that the affine parameter takes the value zero at the horizon of the first black hole and the part of the geodesic that lies outside this horizon in the isotropic co-ordinate patch corresponds to $\lambda > 0$.  Since the isotropic co-ordinate $r$ is such that it limits to the value zero as one approaches the horizon of the first black hole, we should impose the following boundary condition for $r(\lambda)$: 
\be \label{5drbound} r(\lambda = 0) = 0.\ee
The  boundary conditions for the angles are 
\ben \label{anglebound1} \theta(\lambda = 0) &=& \Theta, \qquad \dot{\theta}(\lambda = 0) = 0,  \\ 
\label{anglebound2} \phi(\lambda = 0) &=& \Phi, \qquad \dot{\phi}(\lambda = 0) = 0, \een
where $\Theta$ and $\Phi$ are arbitrary constants at this stage.

The equations \eqref{5drgeod}-\eqref{5dnull} are highly non-linear coupled equations and are probably impossibly to solve directly. The strategy adopted \cite{Candlish1}  is to assume a series expansion for each of the unknown functions $r(\lambda)$, $\theta(\lambda)$, $\phi(\lambda)$. The expansion parameter is an appropriate power of the affine parameter $\lambda$ and it can be motivated as follows.  We  compute the behavior of $r(\lambda)$  near the horizon by examining the leading (in $\lambda$) behavior of  the null condition, which is:
\be \dot{r}^2 =  H \quad \Longrightarrow \quad \dot{r}^2 \sim \frac{1}{r^{2}} \quad \Longrightarrow \quad  r(\lambda)^2 \sim \lambda \qquad \Longrightarrow \qquad r(\lambda) \sim \sqrt{\lambda}.\ee
This together with a similar examination of the behavior of the $\theta$-geodesic and $\phi$-geodesic equations near the horizon, motivates the following series expansion ansatz : 
\be r(\lambda) = \sum_{n = 0}^\infty c_n \, \left(\sqrt{\lambda} \right)^n,\quad \theta(\lambda) = \sum_{n = 0}^\infty b_n \, \left(\sqrt{\lambda} \right)^n, \quad \phi(\lambda) = \sum_{n = 0}^\infty d_n \, \left(\sqrt{\lambda}\right)^n.\ee

The boundary conditions \eqref{5drbound}, \eqref{anglebound1} and \eqref{anglebound2} then imply that the following co-efficients vanish: \be c_0 = 0, \quad b_1 =  0,\quad b_2 = 0, \quad d_1 =  0, \quad  d_2 = 0. \ee 
We thus have 
\be \label{expansionansatz} r(\lambda) = \sum_{n = 1}^\infty c_n \, \left(\sqrt{\lambda} \right)^n,\quad \theta(\lambda) = \Theta + \sum_{n = 3}^\infty b_n \, \left(\sqrt{\lambda} \right)^n, \quad \phi(\lambda) = \Phi +  \sum_{n = 3}^\infty d_n \, \left(\sqrt{\lambda}\right)^n.\ee
The procedure to obtain the solutions to the geodesic equations \cite{Candlish1} is to plug in the expansions \eqref{expansionansatz} into the geodesic equations, obtain a series expansion of the equations in $\sqrt{\lambda}$ and solve order by order. The following paragraphs provides a sketchy summary of implementing this procedure. 

We first examine the $\sqrt{\lambda}$-expansion of the (left hand side of the) null condition \eqref{5dnull}, which starts at ${\cal O}(\lambda^{-1})$. Equating the term at this order to zero determines $c_1$. Successively requiring the vanishing of the terms from ${\cal O}(\lambda^{-1/2})$ to ${\cal O}(\lambda^{3/2})$ determines  $c_2$ to $c_6$. Until this order  one does not encounter any of the $b$'s or $d$'s.  It also turns out that the lowest order at which a certain co-efficient shows up, it shows up linearly.  It does show up non-linearly at higher orders, but by then, it has been determined. Hence at every stage one is solving linear equations.  Then we examine the $\sqrt{\lambda}$-expansion of the (left hand side of the) $\theta$-geodesic equation \eqref{5dthetageod}, which starts at ${\cal O}(\lambda^{-1/2})$. Successively requiring the vanishing of the terms from ${\cal O}(\lambda^{-1/2})$ to ${\cal O}(\lambda^{1/2})$ determines  $b_3$ to $b_5$ in the aforementioned linear way. After this, successively requiring the vanishing of the terms from ${\cal O}(\lambda^{-1/2})$ to ${\cal O}(\lambda^{1/2})$ of the $\phi$-geodesic equation \eqref{5dphigeod} determines  $d_3$ to $d_5$. Then we go back to the null condition, the terms from ${\cal O}(\lambda^{2})$ to ${\cal O}(\lambda^{3})$ determine $c_7$ to $c_9$. Further, the terms from ${\cal O}(\lambda^{1})$ to ${\cal O}(\lambda^{2})$ of the $\theta$-geodesic equation determines $b_6$ to $b_8$, followed by terms from ${\cal O}(\lambda^{1})$ to ${\cal O}(\lambda^{5/2})$ of the $\phi$-geodesic equation determining $d_6$ to $d_9$. Now we can go to ${\cal O}(\lambda^{5/2})$ term of the $\theta$-geodesic equation and determine $b_9$, following which the terms from ${\cal O}(\lambda^{7/2})$ to ${\cal O}(\lambda^{9/2})$ of the null condition allows us to determine the co-efficients $c_{10}$ to $c_{12}$. And so on $\ldots$.  We can continue this process and solve for the expansions of $r(\lambda)$, $\theta(\lambda)$ and $\phi(\lambda)$ to whatever desired order. It is indeed remarkable that these coupled set of highly non-linear equations can be solved using series expansions in a manner reminiscent of the Frobenius method for linear equations.   Here, we give the first few terms,
\ben \label{5drb} r(\lambda) &=& \sqrt{2}\mu_1^{1/4}\lambda^{1/2} + \frac{1}{2\sqrt{2}\mu_1^{1/4}}{\cal G}_0\,\lambda^{\frac32} + \frac25\,{\cal G}_1(\Theta, \Phi)\lambda^2 + \ldots \\ \label{5dthetab} \theta(\lambda) &=& \Theta + \frac{\sqrt{2}}{\mu_1^{1/4}}\partial_\Theta {\cal G}_1(\Theta, \Phi) \lambda^{\frac32} + \frac34 \partial_\Theta {\cal G}_2(\Theta, \Phi) \lambda^2  + \dots \\ \label{5dphib} \phi(\lambda) &=& \Phi +  \frac{\sqrt{2}}{\mu_1^{\frac14}}\,\csc^2 \Theta \,\partial_\Phi {\cal G}_1(\Theta, \Phi)\,\lambda^{\frac32}  + \frac34\,\csc^2 \Theta\,\partial_\Phi {\cal G}_2(\Theta, \Phi)\,\lambda^2  + \ldots, \een
more terms can be found in \eqref{5drsoln}, \eqref{5dthetasoln} and \eqref{5dphisoln} of appendix.  $t(\lambda)$ is then obtained from \eqref{5dtgeod}
\be \label{5dtb} t(\lambda)  \equiv v - T(\lambda, \Theta, \Phi) 
= v + \frac{\mu_1}{4}\,\lambda^{-1} - \frac{3\mu_1^{1/2}}{4} \, {\cal G}_0\, \log \lambda - \frac{8 \sqrt{2}\mu _1^{3/4}}{5}  \,  {\cal G}_1(\Theta, \Phi) \lambda^{1/2} + \ldots   \ee
more terms for  $T(\lambda, \Theta, \Phi)$ can be found in \eqref{5dT} of the appendix. 

We have thus obtained the solutions to the geodesic equations in \eqref{5drsoln}, \eqref{5dthetasoln}, \eqref{5dphisoln} and \eqref{5dT} together with \eqref{5dpsigeod}. As promised in the introduction, the answer is written in terms of the generalized Gegenbauer polynomials ${\cal G}_n(\Theta, \Phi)$ and their derivatives. The expressions are very brief in comparison to analogous expressions written without the aid of the ${\cal G}_n(\Theta, \Phi)$'s. The size of the formulae in \eqref{5drsoln}, \eqref{5dthetasoln} and \eqref{5dT} is the same as those of analogous collinear case formulae; see equations $63, 64$ and $65$ of appendix A of \cite{Candlish1}.

\subsubsection{\label{212}Gaussian null co-ordinates}
We will not give the full theory of Gaussian null co-ordinates here. For this, we refer, apart from  the original reference \cite{Friedrich:1998wq}, to \cite{Candlish1} for a good summary.  We will only note some salient points needed to make sense of subsequent computations. 

The Gaussian null co-ordinate system is a co-ordinate system  well-suited for describing the neighborhood of the horizon (but not so good for describing the asymptotic region). The horizon of a black hole is a codimension  one null hypersurface. Every point in the neighborhood of the horizon is on a single geodesic through some point on the horizon hypersurface. The Gaussian null co-ordinate system  assigns to each point in the neighborhood of the horizon, one co-ordinate corresponding to the value of the affine parameter it takes on the geodesic it lies on, and $d-1$ co-ordinates corresponding to the $d-1$ co-ordinates of the  starting point of the geodesic (in some previously chosen co-ordinate system for the horizon hypersurface). 

In the previous sub-section, we have seen that the null geodesics are parameterized by the $d-1 \,(= 4)$ integration constants viz. $v, \Theta, \Phi, \Psi_1$. It then follows that these are nothing but co-ordinates for the horizon hypersurface and together with the affine parameter $\lambda$ form the co-ordinates of the Gaussian null co-ordinate system. We thus have that $v, \lambda, \Theta, \Phi, \Psi_1$ are the Gaussian null co-ordinates. Now, the  solutions to the geodesic equations are meant to be thought of as providing transition functions from the Gaussian null co-ordinate system to the isotropic co-ordinate system:
\ben \label{tft} t(v, \lambda, \Theta, \Phi, \Psi_1) &=&  v  +  \frac{\mu_1}{4}\lambda^{-1} - \frac{3\mu_1^{1/2}}{4}{\cal G}_0\, \log \lambda - \frac{8 \sqrt{2}\mu _1^{3/4}}{5}  \,  {\cal G}_1(\Theta, \Phi) \lambda^{1/2} + \ldots  \nonumber \\  
\label{tfr} r(v, \lambda, \Theta, \Phi, \Psi_1) &=& \sqrt{2}\mu_1^{1/4}\lambda^{1/2} + \frac{1}{2\sqrt{2}\mu_1^{1/4}}{\cal G}_0\,\lambda^{\frac32} + \frac25\,{\cal G}_1(\Theta, \Phi)\lambda^2 + \ldots \nonumber \\
\label{tftheta} \theta(v, \lambda, \Theta, \Phi, \Psi_1) &=& \Theta + \frac{\sqrt{2}}{\mu_1^{1/4}}\partial_\Theta {\cal G}_1(\Theta, \Phi) \lambda^{\frac32} + \frac34 \partial_\Theta {\cal G}_2(\Theta, \Phi) \lambda^2 + \ldots \nonumber \\
 \label{tfphi} \phi(v, \lambda, \Theta, \Phi, \Psi_1) &=& \Phi +  \frac{\sqrt{2}}{\mu_1^{\frac14}}\,\csc^2 \Theta\,\partial_\Phi {\cal G}_1(\Theta, \Phi)\,\lambda^{\frac32}  + \frac34\,\csc^2 \Theta\,\partial_\Phi {\cal G}_2(\Theta, \Phi)\,\lambda^2 +  \ldots \nonumber  \\
\label{tfpsi} \psi_1(v, \lambda, \Theta, \Phi, \Psi_1) &=& \Psi_1\een

Note that transition functions are not regular at horizon.This is expected because isotropic co-ordinates are not defined at the horizon.
\subsubsection{\label{213}An alternate construction of  Gaussian null co-ordinates}

First we recall few general definitions. A congruence is a family of curves such that through each point  there passes precisely one curve in this family. Tangents to a congruence yield a vector field and conversely, every vector field generates a congruence of curves (whose tangents have the direction of vector field). A vector field is called geodesic if the associated congruence satisfies geodesic equation $k^\beta k_{\alpha;\beta}$. A vector field is twist free (or hypersurface orthogonal) if $k_{\alpha} = \nabla_{\alpha} f$ where $f$ is a scalar. A vector field is null if $k_{\beta}k^{\beta} =0$.  One can easily see that a null, twist-free vector field is automatically geodesic: 
\begin{equation}
k^\beta k_{\alpha;\beta} = f_{;\alpha\beta}f^{;\beta}= f_{;\beta\alpha}f^{;\beta}= \frac{1}{2}(f_{;\beta}f^{;\beta})_{;\alpha}= \frac{1}{2}(k_{\beta}k^{\beta})_{;\alpha} = 0.
\end{equation}
Gaussian null co-ordinates  involve taking a null geodesic and embedding it into a twist-free null geodesic congruence. They are the co-ordinates adapted to the null geodesic congruence and in them the metric takes the form
\begin{equation}
ds^2= 2 dvd\lambda + A(\lambda,x_i, v)dv^2  + 2h_i (\lambda,v,x^j)dvdx^i  + h_{ij}(\lambda,v,x^k)dx^i dx^j
\end{equation}
Here $i,j,k$ range from $1,2, ...,d-2$ with $d$ the spacetime dimension.. If the metric in original coordinates was static then $V= \frac{\partial}{\partial v}$ would be a Killing vector and metric coefficients would be independent of $v$. Hence
\begin{equation}
ds^2= 2 dvd\lambda + A(\lambda,x_i)dv^2  + 2h_i (\lambda,x^j)dvdx^i  + h_{ij}(\lambda,x^k)dx^i dx^j
\end{equation}
Gaussian null co-ordinates are characterized by the conditions 
\be \label{fixedrow}g_{\lambda v} = 1, \quad g_{\lambda\lambda} = 0, \quad g_{\lambda i} = 0.  \ee These  are $d$ coordinate conditions and any metric can be (at least locally) written in this form. In these coordinates, vector field  $\frac{\partial}{\partial \lambda}$ is automatically null and twist-free. By the previous observation, this vector field is also a geodesic vector field, with $\lambda$ playing the role of affine parameter along the null geodesic integral curves of $\frac{\partial}{\partial_{\lambda}}$. Thus a metric in Gaussian null co-ordinates defines a null geodesic congruence in which there is a unique null geodesic through any point, with $\lambda$ an affine parameter on that geodesic and $v,x^i$ are transverse coordinates labeling the geodesics.

So, we can start with the metric in given co-ordinate system and go to Gaussian null co-ordinates in two equivalent ways: 1) use geodesic equations in original metric as done in the bulk of the  paper or 2) use the co-ordinate conditions \eqref{fixedrow}. We apply the second method  to the five dimensional  two centered black hole case, for simplicity (three centered case is quite similar). Consider the  following coordinate transformations
\begin{equation}
t = v - T(\lambda,\Theta) \ \ , \ \  r = g(\lambda,\Theta) \ \ , \ \  \theta = h(\lambda,\Theta)
\end{equation}
and then apply the   co-ordinate conditions \eqref{fixedrow}. These conditions give
\begin{equation}
\partial_{\lambda}T = H^2, \quad  H = (\partial_{\lambda} g)^2 + g^2 (\partial_{\lambda}h)^2, \quad  H\,g^2 \,\partial_{\lambda}h\,\partial_{\Theta}h + H\,\partial_{\lambda}g\,\partial_{\Theta}g = \frac{\partial_{\lambda}T\,\partial_{\Theta}T}{H^2}.
\end{equation}
These look simpler than second order geodesic equations, yet the only way to solve them is through series expansions. We have explicitly checked that they give same solution for $g,h,T$ as do the geodesic equations.

\subsubsection{\label{214}Tensor components in Gaussian null co-ordinates}

We now have the transition functions between the isotropic co-ordinate system and the Gaussian null co-ordinate system \eqref{tfpsi}; obtained by two alternate routes viz. solving the geodesic equations or by solving the co-ordinate conditions \eqref{fixedrow}.  We can now compute the components of the metric and the gauge field in Gaussian null co-ordinates using tensor transformation law. Before we do that, let us note those components of the metric that we \emph{do not have to compute}. It follows from \eqref{fixedrow} that
\be g_{v\lambda}  = 1, \quad g_{\lambda \lambda }  = 0, \quad g_{\lambda \Theta} = 0, \quad g_{\lambda \Phi} = 0, \quad g_{\lambda \Psi_1} = 0. \ee 
If however  we do compute these components, since we have obtained the  transition functions only up to some order in  $\sqrt{\lambda}$, we will be able to verify that they take the above constant values only up to some (related) order, constraining us to infer that they are only finitely differentiable.  But the theory described in the previous section assures us that these components are indeed smooth functions. 

The rest of the metric components in the Gaussian null co-ordinates are given below, only up to the order required to infer their differentiabilties :
\ben \label{gvv} g_{vv} &=& - \frac{4}{\mu_1}\lambda^2 + \frac{12}{\mu _1^{3/2}}{\cal G}_0\, \lambda^3 +  \frac{64 \sqrt{2} }{5 \mu _1^{5/4}} {\cal G}_1(\Theta, \Phi)\lambda^{7/2}  + \ldots \\
\label{gvt} g_{v\Theta}  &=& -\frac{32 \sqrt{2}}{5 \mu _1^{1/4}}\partial_\Theta \,{\cal G}_1(\Theta, \Phi) \lambda^{5/2}  -\frac{20}{3} \partial_\Theta {\cal G}_2(\Theta ,\Phi ) \lambda^3 + \ldots \\
\label{gvp} g_{v\Phi}  &=& -\frac{32 \sqrt{2}}{5 \mu _1^{1/4}}\partial_\Phi \,{\cal G}_1(\Theta, \Phi) \lambda^{5/2}  -\frac{20}{3} \partial_\Phi {\cal G}_2(\Theta ,\Phi ) \lambda^3 + \ldots \een
\begin{multline} \label{gtt} g_{\Theta\Theta}  = \mu_1 + 2\mu_1^{1/2}\,{\cal G}_0 \,\lambda + \frac12 \left[ 2 \,{\cal G}_0^2  + 8 \mu _1\, {\cal G}_2(\Theta ,\Phi ) + 3 \mu _1 \partial^2_\Theta {\cal G}_2(\Theta ,\Phi )\right] \lambda^2 \\ + \frac{\mu _1^{1/4}}{5 \sqrt{2}} \left[ 40 \mu _1\, {\cal G}_3(\Theta ,\Phi ) + 8 \mu _1\,  \partial^2_\Theta {\cal G}_3(\Theta ,\Phi )\right] \lambda^{5/2} + \ldots \end{multline}
\begin{multline} \label{gvp} g_{\Theta\Phi}   =   -\frac{3 \mu _1}{2} \left[ \cot \Theta\,  \partial_\Phi {\cal G}_2(\Theta ,\Phi )  - \partial^2_{\Theta\Phi}{\cal G}_2(\Theta ,\Phi )\right] \lambda^2 - \frac{\mu_1^{1/4}}{5 \sqrt{2}}  \left[ 8\mu_1\,\cot \Theta\,  \partial_\Phi {\cal G}_3(\Theta ,\Phi ) \right. \\ \left.  - 8\mu_1\, \partial^2_{\Theta\Phi}{\cal G}_3(\Theta ,\Phi )\right]\lambda^{5/2} + \ldots \end{multline}
\begin{multline} \label{gpp} g_{\Phi\Phi}  = \mu_1\,\sin^2\Theta +  2\mu_1^{1/2} \,\sin ^2 \Theta  \, {\cal G}_0 \, \lambda   + \frac12 \left[ 2 \, \sin^2 \Theta\, {\cal G}_0^2 + 8 \mu _1\,\sin^2 \Theta \,  {\cal G}_2(\Theta ,\Phi )  \right. \\ \left. + 3\mu _1\, \sin \Theta \, \cos \Theta\,  \partial_\Theta {\cal G}_2 (\Theta ,\Phi)   +3 \mu _1\, \partial^2_\Phi {\cal G}_2(\Theta ,\Phi ) \right] \lambda^2 + \frac{\mu_1^{1/4}}{5 \sqrt{2}} \left[  40 \mu _1 \, \sin^2 \Theta\, {\cal G}_3(\Theta ,\Phi )  \right. \\ \left.  + 8 \mu_1\, \sin \Theta \, \cos  \Theta \,\partial_\Theta {\cal G}_3(\Theta ,\Phi )  +  8 \mu _1\, \partial^2_\Phi {\cal G}_3(\Theta   ,\Phi ) \right] \lambda^{5/2}  + \ldots \end{multline}
\begin{multline}  \label{gpsps} g_{\Psi_1\Psi_1}  = \mu _1\, \sin ^2 \Theta \, \sin ^2 \Phi + 2 \mu_1^{1/2}\,\sin ^2 \Theta \, \sin ^2 \Phi \,   {\cal G}_0 \,\lambda + \frac{\sin \Phi }{2} \left[ 2 \sin^2 \Theta \, \sin \Phi \, {\cal G}_0^2  \right. \\ \left. +  8\mu_1 \, \sin^2 \Theta \, \sin  \Phi \,   {\cal G}_2(\Theta ,\Phi ) +  3\mu_1\, \sin \Theta \, \cos \Theta \, \sin \Phi \, \partial_\Theta {\cal G}_2(\Theta ,\Phi )  + 3\mu_1 \, \cos \Phi \, \partial_\Phi {\cal G}_2(\Theta ,\Phi )\right] \lambda^2  \\ + \frac{ \mu _1^{1/4}}{5 \sqrt{2}} \left[  40\mu_1\, \sin^2 \Theta \,  \sin \Phi\,   {\cal G}_3(\Theta ,\Phi ) +   8\mu_1\, \sin \Theta\,\cos \Theta\,  \sin \Phi \,  \partial_\Theta {\cal G}_3(\Theta ,\Phi )   + 8\mu_1\,  \cos  \Phi   \, \partial_\Phi {\cal G}_3(\Theta ,\Phi )\right] \lambda^{5/2} + \ldots \end{multline}
The components $g_{v\Psi_1}, g_{\Theta\Psi_1}$and $g_{\Phi\Psi_1}$ vanish and hence are smooth.

We can see that out of the fifteen components of the metric, eight of them are constant functions and hence ${\cal C}^\infty$, one of them is ${\cal C}^3$ and the six others are  ${\cal C}^2$. Hence the three center / coplanar black hole metric is  ${\cal C}^2$, i.e. only twice but not thrice differentiable at the horizon of the first black hole; since there is nothing special about the first black hole, the metric is ${\cal C}^2$ at any of the other horizons as well.

Next, we compute the components of the gauge field in the Gaussian null co-ordinate system, again giving terms only up to the order required to infer their differentiabilities:
 \be A = - \frac{dv}{H} + H\,d\lambda + \frac{1}{H} \frac{\partial T}{\partial \Theta}\,d\Theta  + \frac{1}{H} \frac{\partial T}{\partial \Phi}\,d\Phi \ee
\ben A_v &=& -\frac{2}{\mu _1^{1/2}} \lambda  + \frac{3 }{\mu _1} {\cal G}_0\, \lambda^2 + \frac{16 \sqrt{2} }{5 \mu _1^{3/4}} {\cal G}_1(\Theta ,\Phi )\, \lambda^{5/2} + \ldots  \\ A_\lambda &=& \frac{\mu _1^{1/2}}{2}\lambda^{-1} + \frac{3}{4}{\cal G}_0 + \frac{4\sqrt{2} \mu _1^{1/4}}{5}  {\cal G}_1(\Theta ,\Phi )\lambda^{1/2} + \ldots \\ 
 A_\Theta &=& \frac{16 \sqrt{2} \mu _1^{1/4}}{5}  \partial_\Theta {\cal G}_1(\Theta ,\Phi )\, \lambda^{3/2} + \frac{10 \mu _1^{1/2}}{3}  \partial_\Theta {\cal G}_2(\Theta ,\Phi )\, \lambda^2 + \ldots \\
A_\Phi &=& \frac{16 \sqrt{2} \mu _1^{1/4}}{5}  \partial_\Phi {\cal G}_1(\Theta ,\Phi )\, \lambda^{3/2} + \frac{10 \mu _1^{1/2}}{3}  \partial_\Phi {\cal G}_2(\Theta ,\Phi )\, \lambda^2 + \ldots \een
The component $A_{\Psi_1}$ vanishes and hence is smooth. Of the other four, one of the components is a ${\cal C}^2$ function, two of them are ${\cal C}^1$ and one is only ${\cal C}^0$. We thus conclude that the gauge field is only ${\cal C}^0$, i.e. not even once differentiable at any of the horizons. 

\subsubsection{\label{215}Comparing with the two center /  collinear black hole}

As laid out in the introduction, one of the motivations for the computations described in this section was to see if there is any loss of  smoothness to accompany the loss of symmetry from the two center to the three center case.  It is useful to obtain the  two center smoothness results as a special case of our computations.  

To get to the collinear limit,  we set the $a_i$'s to zero in \eqref{3cenharm}. The $R_i$'s reduce to $b_i$'s and for all $i$ the $f_i(\theta, \phi)$ reduces to $\cos \theta$ \eqref{fR}.   More significantly, the generalized Gegenbauer polynomial ${\cal G}_n(\theta, \Phi)$ reduces to the ordinary Gegenbauer polynomial,  $h_n\,G_n(\cos \theta)$, and is a function of only the first variable, it's dependence on the second variable a constant. In the solution to the coplanar problem, we only need to replace all the ${\cal G}_n(\Theta, \Phi)$ by $h_n\,G_n(\cos \Theta)$, $\partial_\Theta {\cal G}_n(\Theta, \Phi)$ by $h_n\,\partial_\Theta G_n(\cos \Theta)$, $\partial_\Phi {\cal G}_n(\Theta, \Phi)$ by $0$, etc and we would have the collinear solution. Doing this for 
 \eqref{5dphisoln}, at least to the order we have computed, we get the expected $\phi(\lambda) = \Phi$. Making the replacements in \eqref{5drsoln}, \eqref{5dthetasoln} and \eqref{5dT} reproduces the results of \cite{Candlish1}.

We now compute the collinear limit of the metric in the Gaussian null co-ordinates. Clearly the eight constant components continue to be smooth functions even in the collinear limit. In addition, we have two other components which are smooth in the collinear limit viz.  $g_{v\Phi}$ and $g_{\Theta\Phi}$ which vanish.   From \eqref{gvv}, $g_{vv}$ continues to be a ${\cal C}^3$ function in the collinear limit. From \eqref{gvt}, we can see that the metric component $g_{v\Theta}$ continues to be a ${\cal C}^2$ function. Similarly we can ascertain that the metric components $g_{\Theta\Theta}$, $g_{\Phi\Phi}$ and $g_{\Psi_1\Psi_1}$ are also ${\cal C}^2$ functions in the collinear limit. Thus the metric in two center black hole solution is ${\cal C}^2$ at any of its horizons. Amongst the gauge field components, $A_\Phi$ vanishes and hence becomes smooth int he collinear limit, while $A_v$, $A_\lambda$ and $A_\theta$ continue to be ${\cal C}^2$, ${\cal C}^0$ and ${\cal C}^1$ functions respectively. Thus the gauge field in the two center black hole solution is ${\cal C}^0$ at any of its horizons. 

The result of our computations is that  \emph{in five dimensions the degree of smoothness of the three center black hole solution is identical to that of the two center black hole solution.} There is no decrease of the degree of smoothness to accompany the decrease in symmetry.  We can make sharper statements in this regard.  When going from the two center to the three center case, only one of the following three things happen for tensor components in the Gaussian null co-ordinate system:
\newline (\textbf{P1}) Components which were smooth continue to be  smooth. 
\newline (\textbf{P2})  Components which were constant and hence smooth become non-constant with a finite degree of smoothness. But the degree of smoothness is not less than the least degree of smoothness already present in the two centre solution. 
\newline (\textbf{P3})  Components which had a finite degree of smoothness are modified; but the modifications are such that the degree of smoothness is unchanged. 

Two other logically allowed possibilities, which don't seem to be realized in the results,  are as follows. One is the opposite of (\textbf{P2}) i.e. that  components become non-constant with a degree of smoothness less than the least degree of smoothness already present in the two center solution, which would result in the three center solution being less smooth than the collinear one. The second is the  opposite of (\textbf{P3}) which is that components with finite degree of smoothness in the two center solution are modified in a manner that reduces their degree of smoothness; again resulting in the coplanar solution being less smooth than the collinear one.

In the five dimensional case that we have been dealing with  $g_{v\Phi},~g_{\Theta\Phi}$ and $A_{\Phi}$ follow (\textbf{P2})  while $g_{vv}, g_{v\Theta}, g_{\Theta\Theta}, g_{\Phi\Phi}, g_{\Psi_1\Psi_1}, A_v, A_\lambda$ and $A_\Theta$ follow (\textbf{P3}) and the rest (\textbf{P1}).  We will see in the subsequent parts of the paper that in every case viz. six and higher dimensional black holes and membranes,  the three center results are related to the two center results by  (\textbf{P1}), (\textbf{P2}) or (\textbf{P3}) only and hence the degree of smoothness is unchanged.

\subsection{\label{22}$d \geq 6$}
We treat all dimensions bigger than five simultaneously. The problem has already been set up in the previous section and in the beginning of this section. The procedure to determine the degree of smoothness of the three center / coplanar solution has already been laid out in the previous sub-section. We will be brief here. We first solve the geodesic equations. 

\subsubsection{\label{221}Solving the geodesic equations}

Here, each of the  $\frac{\partial}{\partial \psi_i}, \quad i = 1, \ldots (d-4)$ are Killing vector fields of the metric and hence each of  ``$\psi_i$-geodesic'' equations admits a first integral and can be readily solved,
\be \label{6dpsigeod}  \psi_i(\lambda) = \Psi_i, \qquad i = 1, 2, \ldots d-4.\ee
The solution to the ``$t$-geodesic'' equation is identical to the $d=5$ case: 
\be \label{6dtgeod}\frac{d}{d\lambda} \left[ H^{-2} \, \frac{dt}{d\lambda} \right] = 0 ~\Longrightarrow ~ \frac{d}{d\lambda}t(\lambda) = -  H(r(\lambda), \theta(\lambda), \phi(\lambda))^2 ~ \Longrightarrow ~  t(\lambda) = v \,- \int d\lambda \,H(r(\lambda), \theta(\lambda), \phi(\lambda))^2. \ee
Again, $t(\lambda)$ is determined in terms of $r(\lambda), \theta(\lambda)$ and $\phi(\lambda)$ via equation \eqref{6dtgeod}, which are obtained by solving simultaneously the ``$r$-geodesic'' equation

\begin{multline} \label{6drgeod} \ddot{r} - H^{\frac{d-5}{d-3}}\,\partial_r H + \frac{\partial_r H}{(d-3) H}\,\dot{r}^2 - \frac{\partial_r H}{ (d-3)H}\,r^2\dot{\theta}^2 - r\dot{\theta}^2 - \frac{\partial_r H}{(d-3) H}\,r^2 \sin^2\theta\, \dot{\phi}^2 - r \sin^2\theta\, \dot{\phi}^2  \\ + \frac{2\, \partial_\theta H}{(d-3)H} \, \dot{r}\,\dot{\theta} + \frac{2\,\partial_\phi H}{(d-3)H}\,\dot{r}\,\dot{\phi} = 0, \end{multline}
the ``$\theta$-geodesic'' equation
\begin{multline} \label{6dthetageod} \ddot{\theta} - H^{\frac{d-5}{d-3}}\,\frac{\partial_\theta H}{r^2} - \frac{\partial_\theta H}{(d-3) H r^2}\,\dot{r}^2 + \frac{\partial_\theta H}{(d-3) H}\,\dot{\theta}^2  -  \frac{\partial_\theta H}{(d-3) H}\,\sin^2\theta\,\dot{\phi}^2 - \sin\theta \cos \theta\,\dot{\phi}^2 \\ + \frac{2}{r}\,\dot{r} \,\dot{\theta} + \frac{2\,\partial_r H}{(d-3)H}\,\dot{r}\,\dot{\theta} + \frac{2\,\partial_\phi H}{(d-3)H}\,\dot{\theta}\,\dot{\phi} = 0,\end{multline}
and  the ``$\phi$-geodesic'' equation
\begin{multline} \label{6dphigeod} \ddot{\phi} - H^{\frac{d-5}{d-3}}\,\frac{\partial_\phi H}{r^2 \sin^2\theta} - \frac{\partial_\phi H}{(d-3) H r^2 \sin^2\theta}\,\dot{r}^2  - \frac{\partial_\phi H}{(d-3) H  \sin^2\theta}\,\dot{\theta}^2 + \frac{\partial_\phi H}{(d-3) H}\,\dot{\phi}^2 \\ + \frac{2}{r}\,\dot{r} \,\dot{\phi} + \frac{2\,\partial_r H}{(d-3)H}\,\dot{r}\,\dot{\phi} + 2 \cot \theta \,\dot{\theta}\,\dot{\phi} + \frac{2\,\partial_\theta H}{(d-3)H}\,\dot{\theta}\,\dot{\phi} = 0.\end{multline}
Again, it will be convenient to work with the null condition
\be \label{6dnull}- H^{-2}\,\dot{t}^2 + H^{\frac{2}{d-3}}\,\dot{r}^2 + H^{\frac{2}{d-3}} r^2\,\dot{\theta}^2 + H^{\frac{2}{d-3}}r^2\,\sin^2\theta\,\dot{\phi}^2 = 0\ee
The boundary conditions are identical to the $d=5$ case \eqref{5drbound}, \eqref{anglebound1} and \eqref{anglebound2}. To solve the geodesic equations \eqref{6drgeod}, \eqref{6dthetageod}, \eqref{6dphigeod}, again we assume a series expansion for each of the unknown functions $r(\lambda)$, $\theta(\lambda)$, $\phi(\lambda)$. The expansion parameter is an appropriate power of the affine parameter $\lambda$ and is determined as before. Near the horizon, the leading (in $\lambda$) behavior of  the null condition:
\be \dot{r}^2 =  H^{\frac{2(d-4)}{d-3}} \quad \Longrightarrow \quad \dot{r}^{2} \sim \frac{1}{r^{2(d-4)}} \quad \Longrightarrow \quad  r(\lambda)^{d-3} \sim \lambda \qquad \Longrightarrow \qquad r(\lambda) \sim \lambda^{\frac{1}{d-3}}.\ee
Hence we assume the following series expansion ansatz : 
\be r(\lambda) = \sum_{n = 0}^\infty c_n \, \left(\lambda^{\frac{1}{d-3}}\right)^n,\quad \theta(\lambda) = \sum_{n = 0}^\infty b_n \, \left(\lambda^{\frac{1}{d-3}}\right)^n, \quad \phi(\lambda) = \sum_{n = 0}^\infty a_n \, \left(\lambda^{\frac{1}{d-3}}\right)^n.\ee
The boundary conditions \eqref{5drbound}, \eqref{anglebound1} \eqref{anglebound2} now imply  that the following co-efficients vanish: 
\ben c_0 &=& 0, \nonumber \\  b_1 &=&  0,\quad b_2 = 0, \ldots \quad b_{d-3} = 0, \nonumber \\ a_1 &=&  0, \quad  a_2 = 0, \ldots \quad a_{d-3} = 0. \een 
We thus have 
\be r(\lambda) = \sum_{n = 1}^\infty c_n \, \left(\lambda^{\frac{1}{d-3}}\right)^n,\quad \theta(\lambda) = \sum_{n = d-2}^\infty b_n \, \left(\lambda^{\frac{1}{d-3}}\right)^n, \quad \phi(\lambda) = \sum_{n = d-2}^\infty a_n \, \left(\lambda^{\frac{1}{d-3}}\right)^n.\ee
We solve the geodesic equations in a manner similar to \ref{211} i.e. we solve the  
null condition, $\theta$-geodesic and $\phi$-geodesic equations order by order in $\lambda^{\frac{1}{d-3}}$ in a certain order. The steps are similar to the $d=5$ case although here there is an additional complication due to the fact that we are dealing with all $d$ simultaneously. The solution is given here (first few terms) 
\ben \label{6drb} r(\lambda) &=& (d-3)^{1/{d-3}}\, \mu_1^{\frac{d-4}{(d-3)^2}}\, \lambda^{1/{d-3}} + \frac{d-4}{2} \,(d-3)^{\frac{4-d}{d-3}} \mu _1^{-\frac{1}{(d-3)^2}} {\cal G}_0\,(\lambda^{1/{d-3}})^{d-2} + \ldots  \\
\label{6dtb} \theta(\lambda) &=&  \Theta + (d-3)^{\frac{1}{d-3}} \mu _1^{-\frac{1}{(d-3)^2}} \partial_\Theta {\cal G}_1(\Theta ,\Phi )\,(\lambda^{1/{d-3}})^{d-2} + \ldots \\
\label{6dpb} \phi(\lambda) &=&  \Phi +   (d-3)^{\frac{1}{d-3}} \mu _1^{-\frac{1}{(d-3)^2}} \csc^2 \Theta \, \partial_\Phi {\cal G}_1(\Theta ,\Phi )\,(\lambda^{1/{d-3}})^{d-2}  +\ldots \een
and more terms can be found in \eqref{6drsoln}, \eqref{6dthetasoln} and \eqref{6dphisoln} of the appendix.  $t(\lambda)$ is then obtained from \eqref{6dtgeod}
\begin{multline} \label{6dtsoln}t(\lambda)  \equiv v - T(\lambda, \Theta, \Phi) = v + \frac{1}{(d-3)^2}\,\mu _1^{2/{d-3}}\, \lambda^{-1}-\frac{d -2}{(d-3)^2}\,\mu _1^{1/{d-3}}\,  {\cal G}_0(\Theta, \Phi)\,\log \lambda \\ - \frac{2d-2}{2d-5}\,(d-3)^{\frac{1}{d-3}}\,\mu_1^{\frac{2d-7}{(d-3)^2}}\,{\cal G}_1(\Theta, \Phi)\,\lambda^{1/{d-3}} - \frac{d}{2d-4}(d-3)^{\frac{2}{d-3}}\,\mu_1^{\frac{3d-11}{(d-3)^2}}\,{\cal G}_2(\Theta, \Phi)\,(\lambda^{1/{d-3}})^2+\ldots\end{multline}
more terms can be found in \eqref{6dT} of the appendix. 

\subsubsection{\label{222}Gaussian null co-ordinates}
The solutions to the geodesic equations are the  transition functions from the Gaussian null co-ordinate system to the isotropic co-ordinate system. 
\ben \label{6dtft}  t(v, \lambda, \Theta, \Phi, \Psi_1, \ldots, \Psi_{d-4}) &=&  v + \frac{1}{(d-3)^2} \mu _1^{2/{d-3}} \lambda^{-1} +\ldots \nonumber \\
\label{6dtfr} r(v, \lambda, \Theta, \Phi, \Psi_1, \ldots, \Psi_{d-4}) &=& (d-3)^{1/{d-3}}\, \mu_1^{\frac{d-4}{(d-3)^2}}\, \lambda^{1/{d-3}}  + \ldots  \nonumber \\
 \label{6dtftheta} \theta(v, \lambda, \Theta, \Phi, \Psi_1, \ldots, \Psi_{d-4}) &=& \Theta  + (d-3)^{\frac{1}{d-3}} \mu _1^{-\frac{1}{(d-3)^2}} \partial_\Theta {\cal G}_1(\Theta ,\Phi )\,(\lambda^{1/{d-3}})^{d-2} + \ldots \nonumber \\
\label{6dtfphi} \phi(v, \lambda, \Theta, \Phi, \Psi_1, \ldots, \Psi_{d-4}) &=& \Phi  +   (d-3)^{\frac{1}{d-3}} \mu _1^{-\frac{1}{(d-3)^2}} \csc^2 \Theta \, \partial_\Phi {\cal G}_1(\Theta ,\Phi )\,(\lambda^{1/{d-3}})^{d-2}  +\ldots \nonumber \\
\label{6dtfpsi} \psi_i(v, \lambda, \Theta, \Phi, \Psi_1, \ldots, \Psi_{d-4}) &=& \Psi_i\een

\subsubsection{\label{223}Tensor components in Gaussian null co-ordinates}
The following components of the metric in Gaussian null co-ordinates are guaranteed to be smooth by the theory and hence we do  not have to compute them. 
\be \label{281}  g_{v\lambda}  = 1, \quad g_{\lambda \lambda }  = 0, \quad g_{\lambda \Theta} = 0, \quad g_{\lambda \Phi} = 0, \quad g_{\lambda \Psi_1} = 0, \quad \ldots \quad g_{\lambda \Psi_{d-4}} = 0. \ee 
The other components are computed using  \eqref{6dtfpsi} in the tensor transformation law; we give terms only up to the order required to infer their differentiabilties :
\begin{multline} \label{6dgvv} g_{vv} = - (d-3)^2 \mu _1^{-2/{d-3}}\,\lambda^2 +  (d-2)(d-3)^2\, \mu _1^{-3/{d-3}}\, {\cal G}_0\,\lambda^3  \\ + \frac{2(d-1)}{2d-5}\,(d-3)^{\frac{3d-8}{d-3}}\,\mu_1^{\frac{2d-5}{(d-3)^2}}{\cal G}_1(\Theta, \Phi) \,\lambda^{\frac{3d-8}{d-3}} +\dots  \end{multline}
\ben  \label{6dgvt} g_{v\Theta}  &=&  \frac{2 (d-1)}{2 d-5}\, (d-3)^{\frac{2 d-5}{d-3}} \, \mu _1^{-\frac{1}{(d-3)^2}} \partial_\Theta {\cal G}_1(\Theta ,\Phi ) \,\lambda ^{\frac{2 d-5}{d-3}} + \ldots \\ \label{6dgvp} g_{v\Phi}  &=& \frac{2 (d-1)}{2 d-5}\, (d-3)^{\frac{2 d-5}{d-3}} \, \mu _1^{-\frac{1}{(d-3)^2}} \partial_\Phi {\cal G}_1(\Theta ,\Phi ) \,\lambda ^{\frac{2 d-5}{d-3}}  + \ldots  \een
\begin{multline} \label{6dgtt} g_{\Theta\Theta} =  \mu _1^{\frac{2}{d-3}} + 2 \, \mu _1^{\frac{1}{d-3}} {\cal G}_0\,\lambda  + \frac{d-2}{d-1} (d-3)^{\frac{2}{d-3}} \mu _1^{\frac{3 d-11}{(d-3)^2}} \left[ \partial^2_\Theta {\cal G}_2(\Theta, \Phi) + \frac{2d-2}{d-2} {\cal G}_2(\Theta, \Phi)\right] \lambda^{\frac{d-1}{d-3}}  + \ldots \end{multline}
\begin{equation} \label{6dgtp} g_{\Theta\Phi}  =  \frac{d-2}{d-1}\,(d-3)^{\frac{2}{d-3}} \mu _1^{\frac{3 d-11}{(d-3)^2}} \left[ \partial^2_{\Theta\Phi}{\cal G}_2(\Theta, \Phi) - \cot \Theta \, \partial_\Phi {\cal G}_2(\Theta, \Phi)\right]\lambda^{\frac{d-1}{d-3}}  + \ldots \end{equation}
\begin{multline} \label{6dgpp} g_{\Phi\Phi} =   \mu _1^{\frac{2}{d-3}} \sin ^2\Theta + 2 \,\mu _1^{\frac{1}{d-3}}\,\sin ^2\Theta  \, {\cal G}_0\, \lambda  + \frac{d-2}{d-1}\,(d-3)^{\frac{2}{d-3}} \mu _1^{\frac{3 d-11}{(d-3)^2}} \left[\partial^2_\Phi {\cal G}_2(\Theta ,\Phi)  \right. \\  \left.  + \sin \Theta \,  \cos \Theta \, \partial_\Theta {\cal G}_2(\Theta ,\Phi ) + \frac{2 d - 2}{d-2}\,\sin ^2 \Theta\, {\cal G}_2(\Theta ,\Phi )  \right] \lambda^{\frac{d-1}{d-3}}  + \ldots \end{multline}
\begin{multline} \label{6dgpsipsi} g_{\Psi_1\Psi_1} =  \mu _1^{\frac{2}{d-3}}\,\sin ^2\Theta \sin ^2\Phi + 2 \,\mu _1^{\frac{1}{d-3}} \sin^2 \Theta\,\sin^2 \Phi \, {\cal G}_0\,\lambda \\ +  \frac{d-2}{d-1}\,(d-3)^{\frac{2}{d-3}} \mu _1^{\frac{3 d-11}{(d-3)^2}} \left[ \sin \Theta \, \cos \Theta \, \sin^2 \Phi \, \partial_\Theta {\cal G}_2(\Theta, \Phi) + \sin \Phi \, \cos \Phi \, \partial_\Phi \, {\cal G}_2(\Theta, \Phi)  \right. \\ \left. + \frac{2\,d - 2}{d-2}\sin^2 \Theta \, \sin^2 \Phi \, {\cal G}_2(\Theta, \Phi)\right] \lambda^{\frac{d-1}{d-3}}+\ldots\end{multline}
\be \label{6dgpsipsi2} g_{\Psi_1\Psi_1} = \frac{g_{\Psi_2\Psi_2}}{\sin^2 \Psi_1}= \frac{g_{\Psi_3\Psi_3}}{\sin^2 \Psi_1\,\sin^2 \Psi_2} = \ldots  = \frac{g_{\Psi_{d-4}\Psi_{d-4}}}{\sin^2 \Psi_1\,\sin^2 \Psi_2 \,\ldots \,\sin^2 \Psi_{d-5}}\ee
The components $g_{v\Psi_i}, g_{\Theta\Psi_i}, g_{\Phi\Psi_i}, g_{\Psi_i\Psi_j}$ vanish and hence are smooth for all $i \neq j \in 1, \ldots, (d-4)$. 

We can see that out of the $\frac{d(d+1)}{2}$ components of the metric, $\frac{d^2-d-4}{2}$ of them are constant functions and hence ${\cal C}^\infty$, one of them viz. $g_{vv}$ \eqref{6dgvv} is ${\cal C}^3$, two of them viz. $g_{v\Theta}, g_{v\Phi}$ \eqref{6dgvt} \eqref{6dgvp}  ${\cal C}^2$ and the rest $d-1$ of them \eqref{6dgtt} - \eqref{6dgpsipsi2} are ${\cal C}^1$ functions. Hence the three center / coplanar black hole metric is  ${\cal C}^1$, i.e. only once but not twice differentiable  at any of the horizons, in $d \geq 6$.

Next, we compute the components of the gauge field in the Gaussian null co-ordinate system:
 \be A = - \frac{dv}{H} + H\,d\lambda + \frac{1}{H} \frac{\partial T}{\partial \Theta}\,d\Theta  + \frac{1}{H} \frac{\partial T}{\partial \Phi}\,d\Phi \ee
\be \label{6dAv} A_v = -(d-3) \mu _1^{-\frac{1}{d-3}} \lambda + \frac{1}{2} (d-2) (d-3)\mu _1^{-\frac{2}{d-3}}{\cal G}_0\,\lambda^2 +  \frac{d-1}{2 d-5}\,(d-3)^{\frac{2 d-5}{d-3}}\mu _1^{-\frac{d-2}{(d-3)^2}}{\cal G}_1(\Theta ,\Phi )\lambda ^{\frac{2 d-5}{d-3}}  + \ldots  \ee
\ben \label{6dAlambda} A_\lambda &=& \frac{\mu _1^{\frac{1}{d-3}}}{(d-3)}\lambda^{-1} +  \frac{d-2}{2 (d-3)}{\cal G}_0 + \frac{d-1}{2 d-5}(d-3)^{\frac{1}{d-3}}\mu _1^{\frac{d-4}{(d-3)^2}}{\cal G}_1(\Theta ,\Phi )\lambda ^{\frac{1}{d-3}}  + \ldots   \\
\label{6dAt} A_\Theta &=& \frac{2 (d-1)}{2 d-5}\,(d-3)^{\frac{d-2}{d-3}}\,\mu _1^{\frac{d-4}{(d-3)^2}}\,\partial_\Theta {\cal G}_1(\Theta ,\Phi )\,\lambda ^{\frac{d-2}{d-3}} + \ldots  \\
 \label{6dAp} A_\Phi  &=& \frac{2 (d-1)}{2 d-5}\,(d-3)^{\frac{d-2}{d-3}}\,\mu _1^{\frac{d-4}{(d-3)^2}}\,\partial_\Phi {\cal G}_1(\Theta ,\Phi )\,\lambda ^{\frac{d-2}{d-3}}  + \ldots  \een
The components $A_{\Psi_i}$ all vanish and hence are smooth. Of the other four, $A_v$  is a ${\cal C}^2$ function, $A_\Theta$ and $A_\Phi$ are  ${\cal C}^1$ with $A_\lambda$  only ${\cal C}^0$. We thus conclude that the gauge field is only ${\cal C}^0$, i.e. not even once differentiable at any of the horizons. 

\subsubsection{\label{224}Comparing with the two center /  collinear black hole}

First, we will rederive the two center / collinear results of \cite{Candlish1} from our formulae. Following the same steps as in \ref{215};  the $\frac{d^2-d-4}{2}$ constant components continue to be smooth functions  in the collinear limit. In addition, $g_{v\Phi}$ and $g_{\Theta\Phi}$ vanish and hence become smooth in the collinear limit.  From \eqref{6dgvv}, $g_{vv}$ continues to be a ${\cal C}^3$ function in the collinear limit. From \eqref{6dgvt}, we can see that the metric component $g_{v\Theta}$ continues to be a ${\cal C}^2$ function. From \eqref{6dgtt} - \eqref{6dgpsipsi2}, we can infer that the metric components $g_{\Theta\Theta}$, $g_{\Phi\Phi}$ and all the $g_{\Psi_i\Psi_i}$'s are ${\cal C}^1$ functions in the collinear limit. Thus the metric in two center black hole solution is ${\cal C}^1$ at any of its horizons. Amongst the gauge field components, $A_\Phi$ vanishes and hence becomes smooth in the collinear limit, while $A_v$, $A_\lambda$ and $A_\theta$ continue to be ${\cal C}^2$, ${\cal C}^0$ and ${\cal C}^1$ functions respectively. Thus the gauge field in the two center black hole solution is ${\cal C}^0$ at any of its horizons. 

Thus the result of our computations is that  even \emph{in six and higher dimensions the degree of smoothness of the three center black hole solution is identical to that of the two center black hole solution.} There is no decrease of the degree of smoothness to accompany the decrease in symmetry.  Furthermore,  when going from the two center to the three center case, the tensor components in the Gaussian null co-ordinate system  behave in a manner similar to the $d=5$ case, i.e. follow only  the three possibilities given in \ref{215}.  $g_{v\Phi},~g_{\Theta\Phi}$ and $A_{\Phi}$ follow (\textbf{P2})  while $g_{vv}, g_{v\Theta}, g_{\Theta\Theta}, g_{\Phi\Phi}, g_{\Psi_i\Psi_i}, A_v, A_\lambda$ and $A_\Theta$ follow (\textbf{P3}) and the rest (\textbf{P1}).

\section{\label{3}Three center / Coplanar $M2$ Branes}

The multi center $M2$ brane solutions we investigate are (bosonic) solutions to eleven dimensional supergravity. Following is the first presentation of the $M2$ brane solution in  isotropic co-ordinates:
\be \label{m2soln}
ds^2 =  H^{-2/3}\,( -dt^2 + dx^2 + dy^2) + H^{1/3}\, ds^2_{\mathbf{R}^8}, \qquad  C_3 = \frac{dt}{H}\ee
where $ds^2_{\mathbf{R}^8}$ is the flat metric of the transverse Euclidean space $\mathbf{R}^8$. $H$ is a harmonic function in the transverse Euclidean space:
\be \label{m2harm}H(\vec{r}) = 1 + \sum_{i = 1}^\infty \frac{\mu_i}{\| \vec{r} - \vec{r}_i \|^{6}}\ee 
$\vec{r}_i$ are points in the transverse Euclidean space which  correspond to the locations of the horizons of the various $M2$ branes and $\| \|$ is the Euclidean norm. We will need to introduce a co-ordinate system for the transverse Euclidean space;  we only need to set $d = 9$ in \eqref{spcoordinates} and in \eqref{flatmetric}. Thus, the co-ordinates in the isotropic co-ordinate system are $t, x, y, r, \theta, \phi, \psi_1, \dots \psi_{5}$.

The harmonic function for the coplanar/ three centre $M2$-brane solution is given in \eqref{3cenharm} with $d = 9$. 
The $M2$-branes are distributed in the $x_1 - x_2$ plane. The first $M2$-brane, the one with charge $\mu_1$ is at the origin in the transverse Euclidean space and for whose horizon we construct a horizon  co-ordinate system. The other $M2$-branes' centers  are located at $(b_i, a_i, 0, \ldots , 0), \quad i = 2, 3, \ldots$ in the transverse Euclidean space. We again re-organize the harmonic function in terms of generalized Gegenbauer polynomials. We only need to set $d=9$ wherever $d$ appears in the  all the formulae from \eqref{3cenharm} to \eqref{3cenharm3}.

 To construct a co-ordinate system for the neighborhood of the horizon of the first $M2$-brane, we will need to solve for radial null geodesics entering that horizon.  

\subsubsection{\label{301}Solving the geodesic equations}
Since, each of the  $\frac{\partial}{\partial \psi_i}, \quad i = 1, \ldots 5$ are Killing vector fields of the metric we can readily solve the  ``$\psi_i$-geodesic'' equations,
\be \label{11dpsigeod}  \psi_i(\lambda) = \Psi_i, \qquad i = 1, 2, \ldots 5.\ee
$\frac{\partial}{\partial t}, \frac{\partial}{\partial x}$ and $\frac{\partial}{\partial y}$ are Killing vector fields of the metric, due to which the  $t$-geodesic, $x$-geodesic and $y$-geodesic equations can be integrated once: 
\be \label{11dtxygeod} \frac{d}{d\lambda} \left[ H^{-2/3} \, \frac{dt}{d\lambda} \right] = 0, \qquad \frac{d}{d\lambda} \left[ H^{-2/3} \, \frac{dx}{d\lambda} \right] = 0, \qquad \frac{d}{d\lambda} \left[ H^{-2/3} \, \frac{dy}{d\lambda} \right] = 0. \ee  
We will solve \eqref{11dtxygeod} in the following way \cite{Gowdigere:2012kq},
\ben \label{txysoln} t(\lambda) &=& v - f(v, X, Y)\,\int d\lambda \, H(r(\lambda), \, \theta(\lambda), \, \phi(\lambda))^{2/3}, \nonumber \\  x(\lambda) &=& X - g(v, X, Y)\,\int d\lambda \, H(r(\lambda), \, \theta(\lambda), \, \phi(\lambda))^{2/3}, \nonumber  \\ y(\lambda) &=& Y - h(v, X, Y)\,\int d\lambda \, H(r(\lambda), \, \theta(\lambda), \, \phi(\lambda))^{2/3}, \een
where $f, g$ and $h$ are arbitrary smooth functions of the  integrations constants $v, X$ and $Y$. We choose to introduce the arbitrary smooth functions $f,g,h$ of integration constants in the above manner because a simple choices such as constant functions or all of them functions of one variable only,  won't make the metric non-singular at the horizon.  It turns out that a completely arbitrary choice of functions $f,g,h$ does not make the metric in these co-ordinates non-singular either. They will need to satisfy various conditions that we will encounter along the way. Although we do not have a solution to all the constraints that the $f,g,h$ would need to satisfy by the end of the analysis, \eqref{fghconstraint1} and \eqref{fghconstraint2}, we do have many examples (see further ahead \eqref{fghexamples}).

Then, we substitute the solutions obtained so far \eqref{11dpsigeod}, \eqref{txysoln} into the remaining geodesic equations: the ``$r$-geodesic'' equation
\begin{multline} \label{11drgeod}\ddot{r} + \frac{\partial_r H}{3 H^{2/3}}(-f^2 + g^2 + h^2) + \frac{\partial_r H}{6 H}\,\dot{r}^2 - \frac{\partial_r H}{6 H}\,r^2\dot{\theta}^2  - r\dot{\theta}^2  \\ - \frac{\partial_r H}{6 H}\,r^2 \sin^2\theta\, \dot{\phi}^2 - r \sin^2\theta\, \dot{\phi}^2   + \frac{\partial_\theta H}{3 H} \, \dot{r}\,\dot{\theta} + \frac{\partial_\phi H}{3 H}\,\dot{r}\,\dot{\phi} = 0,\end{multline}
the ``$\theta$-geodesic'' equation
\begin{multline} \label{11dthetageod}\ddot{\theta} + \frac{\partial_\theta H}{3 H^{2/3} r^2} (-f^2 + g^2 + h^2) - \frac{\partial_\theta H}{6 H r^2}\,\dot{r}^2 + \frac{\partial_\theta H}{6 H}\,\dot{\theta}^2  -  \frac{\partial_\theta H}{6 H}\,\sin^2\theta\,\dot{\phi}^2 \\ - \sin\theta \cos \theta\,\dot{\phi}^2 + \frac{2}{r}\,\dot{r} \,\dot{\theta} + \frac{\partial_r H}{3 H}\,\dot{r}\,\dot{\theta} + \frac{\partial_\phi H}{3 H}\,\dot{\theta}\,\dot{\phi} = 0\end{multline}
and the ``$\phi$-geodesic'' equation
\begin{multline} \label{11dphigeod} \ddot{\phi} + \frac{\partial_\phi H}{3 H^{2/3}r^2 \sin^2\theta} (-f^2 + g^2 + h^2) - \frac{\partial_\phi H}{6 H r^2 \sin^2\theta}\,\dot{r}^2  - \frac{\partial_\phi H}{6 H  \sin^2\theta}\,\dot{\theta}^2 + \frac{\partial_\phi H}{6 H}\,\dot{\phi}^2 \\ + \frac{2}{r}\,\dot{r} \,\dot{\phi} + \frac{\partial_r H}{3 H}\,\dot{r}\,\dot{\phi} + 2 \cot \theta \,\dot{\theta}\,\dot{\phi} + \frac{\partial_\theta H}{3 H}\,\dot{\theta}\,\dot{\phi} = 0.\end{multline}
It is useful to note the null condition 
\be \label{11dnull} H^{2/3}\,(- f^2 + g^2 + h^2 ) + H^{1/3}\,\dot{r}^2 + H^{1/3} r^2\,\dot{\theta}^2 + H^{1/3} r^2\,\sin^2\theta\,\dot{\phi}^2 = 0.\ee
We can now use one of the freedoms in defining the affine parameter to set 
\be \label{fghconstraint1} S ~\equiv~ -f^2 + g^2 + h^2 = -1. \ee
The boundary conditions are identical to the black hole case \eqref{5drbound}, \eqref{anglebound1} and \eqref{anglebound2}. Note that the other freedom in defining the affine parameter has been used in the boundary condition \eqref{5drbound}.

To solve the geodesic equations i.e. \eqref{11drgeod} - \eqref{11dnull} with \eqref{fghconstraint1}, again we assume a series expansion for each of the unknown functions $r(\lambda)$, $\theta(\lambda)$, $\phi(\lambda)$. The expansion parameter is an appropriate power of the affine parameter $\lambda$ and is determined as before. Near the horizon, the leading (in $\lambda$) behavior of  the null condition:
\be \dot{r}^2 =  H^{1/3} \quad \Longrightarrow \quad \dot{r}^{2} \sim \frac{1}{r^{2}} \quad \Longrightarrow \quad  r(\lambda)^2 \sim \lambda \qquad \Longrightarrow \qquad r(\lambda) \sim \sqrt{\lambda}.\ee
Hence we assume the following series expansion ansatz : 
\be r(\lambda) = \sum_{n = 0}^\infty c_n \, \left(\sqrt{\lambda} \right)^n,\quad \theta(\lambda) = \sum_{n = 0}^\infty b_n \, \left(\sqrt{\lambda} \right)^n, \quad \phi(\lambda) = \sum_{n = 0}^\infty d_n \, \left(\sqrt{\lambda}\right)^n.\ee
The boundary conditions \eqref{5drbound}, \eqref{anglebound1} \eqref{anglebound2} now imply  that the following co-efficients vanish: 
\be c_0 = 0, \quad b_1 =  0,\quad b_2 = 0, \quad d_1 =  0, \quad  d_2 = 0. \ee 
We thus have 
\be r(\lambda) = \sum_{n = 1}^\infty c_n \, \left(\sqrt{\lambda} \right)^n,\quad \theta(\lambda) = \Theta + \sum_{n = 3}^\infty b_n \, \left(\sqrt{\lambda} \right)^n, \quad \phi(\lambda) = \Phi +  \sum_{n = 3}^\infty d_n \, \left(\sqrt{\lambda}\right)^n.\ee
We solve the geodesic equations in a manner similar to \ref{211} i.e. we solve the  
null condition, $\theta$-geodesic and $\phi$-geodesic equations order by order in $\sqrt{\lambda}$ in a certain order.  The solution is given here (first few terms) 
\ben \label{11drb} r(\lambda) &=& \sqrt{2}\mu_1^{1/{12}}\lambda^{1/2} + \frac{1}{2\sqrt{2}\mu_1^{5/{12}}}{\cal G}_0\,\lambda^{7/2} + \frac{8}{27 \mu _1^{1/3}}\,{\cal G}_1(\Theta, \Phi)\lambda^4 + \ldots \\ 
\label{11dthetab} \theta(\lambda) &=& \Theta + \frac{4 \sqrt{2}}{35 \mu _1^{5/12}}\partial_\Theta {\cal G}_1(\Theta ,\Phi ) \lambda^{7/2} + \frac{1}{6 \mu _1^{1/3}} \partial_\Theta {\cal G}_2(\Theta, \Phi) \lambda^4 + \ldots  \\ 
\label{11dphib} \phi(\lambda) &=& \Phi + \frac{4 \sqrt{2}}{35 \mu _1^{5/12}}\csc^2\Theta \,\partial_\Phi {\cal G}_1(\Theta ,\Phi ) \lambda^{7/2} + \frac{1}{6 \mu _1^{1/3}} \csc^2\Theta \,\partial_\Phi {\cal G}_2(\Theta, \Phi) \lambda^4 + \ldots \een
and more terms can be found in \eqref{11drsoln}, \eqref{11dthetasoln} and \eqref{11dphisoln} of the appendix.  $t(\lambda), x(\lambda), y(\lambda)$  are obtained from \eqref{txysoln}
\be t(\lambda) = v - f(v, X, Y)\,T(\lambda, \Theta, \Phi), ~ x(\lambda) = X - g(v, X, Y)\,T(\lambda, \Theta, \Phi), ~ y(\lambda) = Y - h(v, X, Y)\,T(\lambda, \Theta, \Phi) \ee
where 
\be T(\lambda, \Theta, \Phi) \equiv \int d\lambda \, H(r(\lambda, \, \theta(\lambda), \, \phi(\lambda))^{2/3}.\ee
The expression for   $T(\lambda, \Theta, \Phi)$ can be found in \eqref{11dT} of the appendix.

\subsubsection{\label{302} Gaussian null-like  co-ordinates}

The solutions to the geodesic equations are the  transition functions from a co-ordinate system for the $M2$ brane horizon  to the isotropic co-ordinate system. We will refer to this horizon co-ordinate system as the Gaussian null-like co-ordinate system. For our purposes the solutions to the geodesic equation provide transition functions to a good co-ordinate system for the horizon, i.e. the metric in these co-ordinates is non-singular. 
\begin{eqnarray} \label{11dtft} t(v, X, Y, \lambda, \Theta, \Phi, \Psi_1, \ldots \Psi_5)  &=& v  -f(v, X, Y) \left[\frac{\mu _1^{1/3}}{4}\lambda^{-1} + \frac{7 }{12 \mu _1^{1/6}}  {\cal G}_0 \lambda^2  + \frac{64 \sqrt{2}}{135 \mu _1^{1/{12}}}  {\cal G}_1(\Theta ,\Phi ) + \ldots \right]\nonumber \\ 
\label{11dtfx} x(v, X, Y, \lambda, \Theta, \Phi, \Psi_1, \ldots \Psi_5) &=& X  -g(v, X, Y) \left[\frac{\mu _1^{1/3}}{4}\lambda^{-1} + \frac{7 }{12 \mu _1^{1/6}}  {\cal G}_0 \lambda^2  + \frac{64 \sqrt{2}}{135 \mu _1^{1/{12}}}  {\cal G}_1(\Theta ,\Phi ) + \ldots \right]\nonumber \\
\label{11dtfy} y(v, X, Y, \lambda, \Theta, \Phi, \Psi_1, \ldots \Psi_5) &=& Y  -h(v, X, Y) \left[\frac{\mu _1^{1/3}}{4}\lambda^{-1} + \frac{7 }{12 \mu _1^{1/6}}  {\cal G}_0 \lambda^2  + \frac{64 \sqrt{2}}{135 \mu _1^{1/{12}}}  {\cal G}_1(\Theta ,\Phi ) + \ldots \right]\nonumber \\
 \label{11dtfr} r(v, X, Y, \lambda, \Theta, \Phi, \Psi_1, \ldots \Psi_5) &=&  \sqrt{2}\mu_1^{1/{12}}\lambda^{1/2} + \frac{1}{2\sqrt{2}\mu_1^{5/{12}}}{\cal G}_0 \,\lambda^{7/2} + \frac{8}{27 \mu _1^{1/3}}\,{\cal G}_1(\Theta, \Phi)\lambda^4  + \ldots \nonumber \\ 
\label{11dtftheta} \theta(v, X, Y, \lambda, \Theta, \Phi, \Psi_1, \ldots \Psi_5) &=&  \Theta + \frac{4 \sqrt{2}}{35 \mu _1^{5/12}}\partial_\Theta {\cal G}_1(\Theta ,\Phi ) \lambda^{7/2} + \frac{1}{6 \mu _1^{1/3}} \partial_\Theta {\cal G}_2(\Theta, \Phi) \lambda^4  +  \ldots \nonumber \\
 \label{11dtfphi} \phi(v, X, Y, \lambda, \Theta, \Phi, \Psi_1, \ldots \Psi_5) &=& \Phi + \frac{4 \sqrt{2}}{35 \mu _1^{5/12}}\csc^2\Theta \,\partial_\Phi {\cal G}_1(\Theta ,\Phi ) \lambda^{7/2} + \frac{1}{6 \mu _1^{1/3}} \csc^2\Theta \,\partial_\Phi {\cal G}_2(\Theta, \Phi) \lambda^4  + \ldots \nonumber \\
\label{11dtfpsi} \psi_i(v, X, Y, \lambda, \Theta, \Phi, \Psi_1, \ldots \Psi_5) &=& \Psi_i \een

\subsubsection{\label{303} Tensor components in Gaussian null-like co-ordinates}

We can now compute the components of the metric and tensor gauge fields in the Gaussian null-like co-ordinate system using \eqref{11dtfpsi} in the tensor transformation law.  Before we give the explicit formulae, we note that some components are guaranteed to be constant and hence smooth even before doing to the series computations. Consider the following components and their tensor transformation law
\be  g_{\lambda \lambda} = \frac{\partial x^\mu}{\partial \lambda}\,\frac{\partial x^\nu}{\partial \lambda}\,g_{\mu\nu}(x)~, \quad g_{\lambda y} = \frac{\partial x^\mu}{\partial \lambda}\,\frac{\partial x^\nu}{\partial y}\,g_{\mu\nu}(x) \ee 
where $x^\mu$ are isotropic co-ordinates and $y$ is any Gaussian null-like co-ordinate other than $\lambda$. The expression for $g_{\lambda \lambda}$ is nothing but the null condition and hence vanishes. 
Taking derivative w.r.t $\lambda$ of $g_{\lambda y}$ gives
\ben \frac{\partial}{\partial \lambda}g_{\lambda y} &=& \frac{\partial^2 x^\mu}{\partial \lambda^2}\,\frac{\partial x^\nu}{\partial y}\,g_{\mu\nu} + \frac{\partial x^\mu}{\partial \lambda}\,\frac{\partial^2 x^\nu}{\partial \lambda \partial y}\,g_{\mu\nu}  + \frac{\partial x^\mu}{\partial \lambda}\,\frac{\partial x^\nu}{\partial y}\,\frac{\partial x^\rho}{\partial \lambda}\,\partial_\rho g_{\mu\nu} \nonumber \\  &=& - \Gamma^\mu_{\rho \sigma}\,\frac{\partial x^\rho}{\partial \lambda} \,\frac{\partial x^\sigma}{\partial \lambda}\, \frac{\partial x^\nu}{\partial y}\,g_{\mu\nu}  - \frac12 \frac{\partial x^\mu}{\partial \lambda}\,\frac{\partial x^\nu}{\partial \lambda }\, \frac{\partial x^\rho}{\partial y }\partial_\rho g_{\mu\nu}  + \frac{\partial x^\mu}{\partial \lambda}\,\frac{\partial x^\nu}{\partial y}\,\frac{\partial x^\rho}{\partial \lambda}\,\partial_\rho g_{\mu\nu}.  \een
We have used  the geodesic equation in the first term, for the second term we have used the $\frac{\partial}{\partial y}$ of the null condition. Simplifying, we find :
\be \frac{\partial}{\partial \lambda}\,g_{\lambda y} = 0. \ee
This means that in the series expansion of these components of the metric, there is only one term, the $(\sqrt{\lambda})^0$ term, which we can readily compute; more importantly these components of the metric are all smooth functions. We thus have the following smooth components:
\be \label{3126} g_{\lambda \lambda} = 0, \quad g_{\lambda v} = f, \quad g_{\lambda X} = -g, \quad g_{\lambda Y} = - h, \quad g_{\lambda \Theta} = 0, \quad g_{\lambda \Phi} = 0, \quad g_{\lambda \Psi_i} = 0. \ee
The other non-vanishing components of the metric in the  Gaussian null-like co-ordinates are given below.
\ben \label{11dgvv} g_{vv} &=&  \frac{1}{4} \mu _1^{1/3}  z_1 - 2\,  \partial_vf\,\lambda - \frac{4}{\mu _1^{1/3}} \lambda^2 -\frac{7}{3 \mu _1^{1/6}}  z_1 {\cal G}_0\, \lambda^3 -\frac{32 \sqrt{2} }{15 \mu _1^{1/{12}}} z_1 {\cal G}_1(\Theta ,\Phi ) \lambda^{7/2}  + \ldots \nonumber 
\\
 \label{11dgXX} g_{XX} &=&  \frac{1}{4} \mu _1^{1/3}  z_2 + 2\,  \partial_Xg\,\lambda + \frac{4}{\mu _1^{1/3}}\lambda^2 -\frac{7}{3 \mu _1^{1/6}}  z_2  {\cal G}_0\, \lambda^3 -\frac{32 \sqrt{2} }{15 \mu _1^{1/{12}}} z_2 {\cal G}_1(\Theta ,\Phi ) \lambda^{7/2}  + \ldots \nonumber \\
\label{11dgYY} g_{YY} &=&  \frac{1}{4} \mu _1^{1/3}  z_3 + 2\,  \partial_Yh\,\lambda + \frac{4}{\mu _1^{1/3}}\,\lambda^2 -\frac{7}{3 \mu _1^{1/6}}  z_3\,{\cal G}_0\, \lambda^3 -\frac{32 \sqrt{2} }{15 \mu _1^{1/{12}}} z_3\, {\cal G}_1(\Theta ,\Phi ) \lambda^{7/2}  + \ldots \nonumber \\
 \label{11dgvX} g_{vX} &=&  \frac{1}{4} \mu _1^{1/3}  q_2 -  q_1\, \lambda -\frac{7}{3 \mu _1^{1/6}}  q_2\, {\cal G}_0\, \lambda^3 -\frac{32 \sqrt{2} }{15 \mu _1^{1/{12}}} q_2\, {\cal G}_1(\Theta ,\Phi ) \lambda^{7/2}  + \ldots \nonumber \\ 
 \label{11dgvY} g_{vY} &=&  \frac{1}{4} \mu _1^{1/3}  q_4 -  q_3\, \lambda -\frac{7}{3 \mu _1^{1/6}}  q_4 \, {\cal G}_0\, \lambda^3 -\frac{32 \sqrt{2} }{15 \mu _1^{1/{12}}} q_4\, {\cal G}_1(\Theta ,\Phi ) \lambda^{7/2}  + \ldots \nonumber \\ \vdots \nonumber \een
\ben \vdots \nonumber \\ \label{11dgXY} g_{XY} &=&  \frac{1}{4} \mu _1^{1/3}  q_6  -  q_5\, \lambda -\frac{7}{3 \mu _1^{1/6}}  q_6 \, {\cal G}_0\, \lambda^3 -\frac{32 \sqrt{2} }{15 \mu _1^{1/{12}}} q_6\, {\cal G}_1(\Theta ,\Phi ) \lambda^{7/2} + \ldots  \een
\ben \label{11dgvtheta} \frac{g_{v\Theta}}{f}  &=& - \frac{g_{X\Theta}}{g} = - \frac{g_{Y\Theta}}{h}  =  \frac{256 \sqrt{2}}{135 \mu _1^{5/12}} \, \partial_\Theta {\cal G}_1(\Theta ,\Phi )\lambda^{9/2} + \frac{16}{5 \mu _1^{1/3}} \,\partial_\Theta {\cal G}_2(\Theta ,\Phi )\lambda^5  + \ldots \\ \label{11dgvphi} \frac{g_{v\Phi}}{f} &=& - \frac{g_{X\Phi}}{g} = - \frac{g_{Y\Phi}}{h}  =  \frac{256 \sqrt{2}}{135 \mu _1^{5/12}} \, \partial_\Phi {\cal G}_1(\Theta ,\Phi )\lambda^{9/2} + \frac{16}{5 \mu _1^{1/3}} \,\partial_\Phi {\cal G}_2(\Theta ,\Phi )\lambda^5  + \ldots \een
\be \label{11dgtt} g_{\Theta\Theta} = \mu_1^{1/3} + \frac{8 }{3 \mu _1^{1/6}} {\cal G}_0\,\lambda^3+ \frac{256 \sqrt{2}}{105 \mu _1^{1/{12}}} {\cal G}_1(\Theta ,\Phi ) \lambda^{7/2}+ \frac13 \left[ 16 {\cal G}_2(\Theta ,\Phi )+ \partial^2_\Theta {\cal G}_2(\Theta ,\Phi )\right]\lambda^4  + \ldots \ee
\be \label{11dgtp} g_{\Theta\Phi} = \frac13 \left[ \partial^2_{\Theta\Phi} {\cal G}_2(\Theta ,\Phi ) - \cot  \Theta \, \partial_\Phi {\cal G}_2(\Theta ,\Phi ) \right] \lambda^4 +\frac{16\sqrt{2} \mu _1^{1/{12}}}{63}\left[ \partial^2_{\Theta\Phi} {\cal G}_3(\Theta ,\Phi ) - \cot  \Theta \, \partial_\Phi {\cal G}_3(\Theta ,\Phi )\right] \lambda^{9/2} +\ldots \ee
\be \label{11dgpp} \frac{g_{\Phi\Phi}}{\sin^2 \Theta} = \mu_1^{1/3} + \frac{8 }{3 \mu _1^{1/6}} \,{\cal G}_0\,\lambda^3 + \frac{8 \sqrt{2}}{105\mu _1^{1/{12}}}  \left[ 3\csc^2 \Theta \,\partial^2_\Phi {\cal G}_1(\Theta ,\Phi )  + 3 \cot \Theta \, \partial_\Theta {\cal G}_1(\Theta ,\Phi )    + 35  {\cal G}_1(\Theta ,\Phi )  \right] \lambda^{7/2}   + \ldots 
\ee
\begin{multline} \label{11dgpsipsi} \frac{g_{\Psi_1\Psi_1}}{\sin^2 \Theta\,\sin^2\Phi} = \mu_1^{1/3} + \frac{8 }{3 \mu _1^{1/6}}  \,{\cal G}_0\,\lambda^3 + \frac{8 \sqrt{2}}{105\mu _1^{1/{12}}}  \left[ 3\, \csc^2 \Theta \,\partial^2_\Phi {\cal G}_1(\Theta ,\Phi )  \right. \\ \left. + 3\, \csc^2 \Theta \, \cot \Phi \, \partial_\Phi {\cal G}_1(\Theta, \Phi) + 3\, \cot \Theta \, \partial_\Theta {\cal G}_1(\Theta ,\Phi )   + 35 \, {\cal G}_1(\Theta ,\Phi )  \right] \lambda^{7/2}  + \ldots 
\end{multline}
\be g_{\Psi_1\Psi_1} = \frac{g_{\Psi_2\Psi_2}}{\sin^2 \Psi_1}= \frac{g_{\Psi_3\Psi_3}}{\sin^2 \Psi_1\,\sin^2 \Psi_2} = \frac{g_{\Psi_4\Psi_4}}{\sin^2 \Psi_1\,\sin^2 \Psi_2 \,\sin^2 \Psi_3} = \frac{g_{\Psi_5\Psi_5}}{\sin^2 \Psi_1\,\sin^2 \Psi_2 \,\sin^2 \Psi_3 \,\sin^2 \Psi_4}\ee
where $z_1$ - $z_3$ and $q_1$ - $q_6$ are the following functions:
\begin{eqnarray} \label{b2}
q_1(v, X, Y) & \equiv &  \partial_X f - \partial_v g, \qquad \qquad    q_2(v, X, Y)  \equiv   -\partial_v f \, \partial_X f+ \partial_v g \, \partial_X g + \partial_v h \, \partial_X h   \nonumber \\
q_3(v, X, Y) & \equiv &  \partial_Y f - \partial_v h, \qquad \qquad q_4(v, X, Y)  \equiv   -\partial_v f \, \partial_Y f+ \partial_v g \, \partial_Y g + \partial_v h \, \partial_Y h \nonumber \\
q_5(v, X, Y) & \equiv & - \left(\partial_X  h + \partial_Y g\right), \qquad  q_6(v, X, Y)  \equiv   -\partial_Y f \, \partial_X f+ \partial_Y g \, \partial_X g + \partial_Y h \, \partial_X h,  \nonumber \\
z_1(v, X, Y) & \equiv &  -\left(\partial_v f \right)^2+\left(\partial_v g \right)^2+\left(\partial_v h \right)^2,  \quad z_2(v, X, Y) \equiv   -\left(\partial_X f \right)^2+\left(\partial_X g \right)^2+\left(\partial_X h \right)^2,  \nonumber \\
z_3(v, X, Y) & \equiv &  -\left(\partial_Y f \right)^2+\left(\partial_Y g \right)^2+\left(\partial_Y h \right)^2.
\end{eqnarray}

Apart from these, we also require that the metric is non-singular at $\lambda = 0$. We can compute the determinant of the metric on the horizon i.e. at $\lambda = 0$:
\ben \label{det} g = \frac{\mu_1^3}{16}  g_{S^7}\, [ f^2 (q_6^2 - z_2 z_3) + g^2 (q_4^2 - z_3 z_1) + h^2 (q_2^2 - z_1 z_2)   \nonumber \\ + 2 f g (q_4 q_6 - q_2 z_3) - 2 g h (q_2 q_4 - q_6 z_1) +  2 f h (q_6 q_2 - q_4 z_2)   ]\een
where $g_{S_7}$ is the determinant of the round metric on the unit seven sphere and  the $q_i$'s and the $z_i$'s are defined in \eqref{b2}. Requiring that the determinant does not vanish on the horizon gives us the following condition that our choice of $f, g, h$ functions must satisfy:
\be  \label{fghconstraint}  f^2 (q_6^2 - z_2 z_3) + g^2 (q_4^2 - z_3 z_1) + h^2 (q_2^2 - z_1 z_2)   + 2 f g (q_4 q_6 - q_2 z_3) - 2 g h (q_2 q_4 - q_6 z_1) +  2 f h (q_6 q_2 - q_4 z_2)  \neq 0. \ee 
In an appendix, we show that this condition that the $f, g, h$ functions need to satisfy reduces to the following succinct condition. We have relegated these manipulations to an appendix in this paper but we expect it to contribute to  the complete theory of Gaussian null co-ordinates for the membrane horizon, which is presently lacking and which we hope to provide in a future work. 
\be \label{fghconstraint2} \nabla g \times \nabla h \neq 0, \ee
where $\nabla g$ is the 3-vector whose components are the $v, X$ and $Y$ derivatives of $g$, similarly $\nabla h$.
The simple choice of constant functions for $f, g, h$ does not satisfy the above constraint. The choice of all of them functions of only one variable also does not satisfy the above constraint. But we do have many examples for the $f, g, h$ functions that satisfy \eqref{fghconstraint1} and \eqref{fghconstraint2} two of which are
\begin{eqnarray} \label{fghexamples}
f(v, X, Y) &=&\frac{1}{2}\left( X + \frac{1}{X} + \frac{Y^2}{X} \right), \quad g(v, X, Y)= \frac{1}{2}\left( - X + \frac{1}{X} + \frac{Y^2}{X} \right), \quad h(v, X, Y)= Y. \nonumber \\
f(v, X, Y)&=&\sqrt{1+ Y^2}\,\cosh X ,\quad g(v, X, Y)= \sqrt{1 + Y^2}\,\sinh X , \quad h(v, X, Y) = Y.
\end{eqnarray}

We can see that out of the $66$ components of the metric in the Gaussian null-like co-ordinate system, $46$ of them are ${\cal C}^\infty$, $7$ of them are ${\cal C}^4$ and the $13$ others are  ${\cal C}^3$. Hence the three center / coplanar $M2$ brane metric is  ${\cal C}^3$, i.e. only thrice but not four times differentiable at the horizon of the first $M2$ brane's horizon and at  any of the other horizons as well.

Next, we give the non-vanishing components of the tensor gauge field in the Gaussian null-like co-ordinate system  to the relevant order needed to infer the differentiability. 
\ben \label{CvXl} 
C_{vX\lambda} &=& \frac{\mu_1^{1/2}}{8}u_4 \,\lambda^{-1} -\frac{\mu_1^{1/6}}{2} u_3 - \frac{2}{\mu_1^{1/6}} h\, \lambda - \frac{7}{8}  u_4\, {\cal G}_0\,\lambda ^2 -\frac{104\sqrt{2}\,\mu_1^{1/{12}}}{135} u_4\,  {\cal G}_1(\Theta ,\Phi )\lambda ^{5/2} + \ldots \nonumber \\
 \label{CvYl} 
C_{vY\lambda} &=& \frac{\mu_1^{1/2}}{8}u_6 \,\lambda^{-1} -\frac{\mu_1^{1/6}}{2} u_5 + \frac{2}{\mu_1^{1/6}} g\,\lambda - \frac{7}{8}  u_6\, {\cal G}_0\,\lambda ^2 -\frac{104\sqrt{2}\,\mu_1^{1/{12}}}{135} u_6\,  {\cal G}_1(\Theta ,\Phi )\lambda ^{5/2} + \ldots \nonumber \\
\label{CXYl}
C_{XY\lambda} &=& \frac{\mu_1^{1/2}}{8}u_8 \,\lambda^{-1} -\frac{\mu_1^{1/6}}{2} u_7 - \frac{2}{\mu_1^{1/6}} f\, \lambda - \frac{7}{8}  u_8\, {\cal G}_0\,\lambda ^2 -\frac{104\sqrt{2}\,\mu_1^{1/{12}}}{135} u_8\,  {\cal G}_1(\Theta ,\Phi )\lambda ^{5/2} + \ldots \nonumber   \\
\een
\ben \label{CvXt} 
C_{vX\Theta} &=& \frac{32\sqrt{2}\mu_1^{1/{12}}}{135} u_4\, \partial_\Theta{\cal G}_1(\Theta ,\Phi)\, \lambda ^{7/2} + \frac{2\mu_1^{1/6}}{5}u_4\,  \partial_\Theta {\cal G}_2(\Theta ,\Phi )\, \lambda ^4 + \ldots \nonumber \\
\label{CvYt} 
C_{vY\Theta} &=& \frac{32\sqrt{2}\mu_1^{1/{12}}}{135} u_6\, \partial_\Theta{\cal G}_1(\Theta ,\Phi)\, \lambda ^{7/2} + \frac{2\mu_1^{1/6}}{5}u_6\,  \partial_\Theta {\cal G}_2(\Theta ,\Phi )\, \lambda ^4 + \ldots \nonumber \\
\label{CXYt} 
C_{XY\Theta} &=& \frac{32\sqrt{2}\mu_1^{1/{12}}}{135} u_8\, \partial_\Theta{\cal G}_1(\Theta ,\Phi)\, \lambda ^{7/2} + \frac{2\mu_1^{1/6}}{5}u_8\,  \partial_\Theta {\cal G}_2(\Theta ,\Phi )\, \lambda ^4 + \ldots \nonumber \\
\label{CvXp} 
C_{vX\Phi} &=& \frac{32\sqrt{2}\mu_1^{1/{12}}}{135} u_4\, \partial_\Phi{\cal G}_1(\Theta ,\Phi)\, \lambda ^{7/2} + \frac{2\mu_1^{1/6}}{5}u_4\,  \partial_\Phi {\cal G}_2(\Theta ,\Phi )\, \lambda ^4 + \ldots \nonumber \\
\label{CvYp} 
C_{vY\Phi} &=& \frac{32\sqrt{2}\mu_1^{1/{12}}}{135} u_6\, \partial_\Phi{\cal G}_1(\Theta ,\Phi)\, \lambda ^{7/2} + \frac{2\mu_1^{1/6}}{5}u_8\,  \partial_\Phi {\cal G}_2(\Theta ,\Phi )\, \lambda ^4 + \ldots \nonumber \\
\label{CXYp} 
C_{XY\Phi} &=& \frac{32\sqrt{2}\mu_1^{1/{12}}}{135} u_8\, \partial_\Phi{\cal G}_1(\Theta ,\Phi)\, \lambda ^{7/2} + \frac{2\mu_1^{1/6}}{5}u_8\,  \partial_\Phi {\cal G}_2(\Theta ,\Phi )\, \lambda ^4 + \ldots 
\een
\be\label{CvXY} 
C_{vXY} = -\frac{\mu_1^{1/6}}{2}u_2\,\lambda -\frac{2}{\mu_1^{1/6}}u_1\,\lambda ^2 -\frac{8}{\mu_1^{1/2}}\lambda ^3+ \frac{35}{6\mu_1^{1/3}}u_2\,{\cal G}_0\,\lambda ^4 + \frac{736 \sqrt{2}}{135 \mu_1^{1/4}}u_2\, {\cal G}_1(\Theta ,\Phi )\, \lambda ^{9/2}  + \ldots
\ee
where
\begin{eqnarray} \label{M2b3}
u_1(v, X, Y) & \equiv & \partial_{Y}h + \partial_v f + \partial_X g, \quad
u_3(v, X, Y)  \equiv  h \, \partial_v f - f \, \partial_v h + h \, \partial_X g -g \, \partial_X h, \nonumber \\
u_5(v, X, Y) & \equiv & f \, \partial_v g - g \, \partial_v f + h \, \partial_{Y} g -g \, \partial_{Y} h, \quad
u_7(v, X, Y)  \equiv  f \, \partial_{Y} h - h \, \partial_{Y} f + f \, \partial_X g -g \, \partial_X f, \nonumber \\
u_2(v, X, Y) & \equiv & \left(\partial_v f \, \partial_{Y}h - \partial_{Y}f \, \partial_v h \right) + \left(\partial_X g \, \partial_{Y}h - \partial_{Y}g \, \partial_X h \right)  + \left(\partial_v f \, \partial_{X}g - \partial_{X}f \, \partial_v g \right), \nonumber 
 \een
\ben 
u_4(v, X, Y) & \equiv & f  \left(\partial_v h \, \partial_X g - \partial_X h \, \partial_v g \right) + g \left(\partial_v f \, \partial_X h - \partial_X f \, \partial_v h \right) + h \left(\partial_v g \, \partial_X f - \partial_X g \, \partial_v f \right), \nonumber \\
u_6(v, X, Y) & \equiv & f  \left(\partial_v h \, \partial_{Y} g - \partial_{Y} h \, \partial_v g \right) + g  \left(\partial_v f \, \partial_{Y} h - \partial_{Y} f \, \partial_v h \right)  + h \left(\partial_v g \, \partial_{Y} f - \partial_{Y} g \, \partial_v f \right), \nonumber \\
u_8(v, X, Y) & \equiv & f  \left(\partial_X h \, \partial_{Y} g - \partial_{Y} h \, \partial_X g \right) + g  \left(\partial_X f \, \partial_{Y} h - \partial_{Y} f \, \partial_X h \right) + h  \left(\partial_X g \, \partial_{Y} f - \partial_{Y} g \, \partial_X f \right). \nonumber
\end{eqnarray}
We observe that out of $165$ components of the tensor gauge field, $155$ of them are are ${\cal C}^\infty$, three of them ${\cal C}^2$ \eqref{CXYl},  six are ${\cal C}^3$ \eqref{CXYp} and  one is ${\cal C}^4$ \eqref{CvXY}. Thus we conclude that the gauge field is ${\cal C}^2$ at any of the horizons.

\subsubsection{\label{304} Comparing with  two center / collinear $M2$ branes}
First, we will rederive the two center / collinear results of \cite{Gowdigere:2012kq} from our formulae. Following the same steps as in \ref{215};  the $46$ smooth components continue to be smooth functions  in the collinear limit. In addition, $g_{v\Phi}, g_{X\Phi}, g_{Y\Phi}$ and $g_{\Theta\Phi}$ vanish and hence become smooth in the collinear limit.  From \eqref{11dgvtheta}, we can see that the metric components $g_{v\Theta}, g_{X\Theta}$ and $g_{Y\Theta}$ are ${\cal C}^4$ functions in the collinear limit. All other metric components are  ${\cal C}^3$ functions in the collinear limit.   Thus the metric in two center $M2$ brane solution is ${\cal C}^3$ at any of its horizons. Amongst the gauge field components, the $155$ smooth components continue to be smooth functions  in the collinear limit. In addition, $C_{vX\Phi}, C_{vY\Phi}$ and $C_{XY\Phi}$ vanish and hence become smooth in the collinear limit.  $C_{vX\lambda}, C_{vY\lambda}$ and $C_{XY\lambda}$ are ${\cal C}^2$ \eqref{CXYl}, $C_{vX\Theta}, C_{vY\Theta}$ and $C_{XY\Theta}$ are ${\cal C}^3$ \eqref{CXYt} and $C_{vXY}$ is ${\cal C}^4$ \eqref{CvXY} in the collinear limit.  Thus the tensor gauge field in two center $M2$ brane solution is ${\cal C}^2$ at any of its horizons.

Thus the result of our computations is that  even \emph{for $M2$ branes, the degree of smoothness of the three center  solution is identical to that of the two center  solution.} There is no decrease of the degree of smoothness to accompany the decrease in symmetry.  Furthermore, when going from the two center to the three center situation, the tensor components in the Gaussian null-like  co-ordinate system  behave in a manner similar to the  black hole case, i.e. follow only  the three possibilities given in \ref{215}.  $g_{v\Phi}, g_{X\Phi}, g_{Y\Phi},  g_{\Theta\Phi}, C_{vX\Phi}, C_{vY\Phi}$ and $C_{XY\Phi}$  follow (\textbf{P2}). The $16$ metric components  $g_{vv}, g_{XX}, g_{YY}, g_{vX}, g_{vY}, g_{YY}, g_{v\Theta}, g_{X\Theta}, g_{Y\Theta}, g_{\Theta\Theta}, g_{\Phi\Phi}, g_{\Psi_i\Psi_i}$ and the $7$ tensor gauge field components $C_{vX\lambda}, C_{vY\lambda}, C_{XY\lambda}, C_{vX\Theta}, C_{vY\Theta},  C_{XY\Theta}, C_{vXY}$ follow (\textbf{P3}) and the rest  follow (\textbf{P1}).  

\section{\label{4} Conclusion and Outlook}
We have  obtained answers to the various questions that motivated this work.  To begin with, we have an answer to the basic question: what is the degree of differentiability  at a horizon of the various tensor fields in the three-center /  coplanar solution? Expressing the harmonic function in terms of generalized Gegenbauer polynomials allows us to do the computations and obtain the results in a compact manner.  For the $d=5$ black hole, the metric is twice but not thrice differentiable (${\cal C}^2$) and the gauge field is not even once differentiable (${\cal C}^0$) at any of the horizons \ref{214}. For the black hole in six and higher dimensions, the metric is once but not twice differentiable (${\cal C}^2$) and the gauge field is not even once differentiable (${\cal C}^0$) at any of the horizons \ref{223}. And for the $M2$ branes, the metric is thrice but not four times differentiable (${\cal C}^3$) and the gauge field is twice but not thrice  differentiable (${\cal C}^2$) at any of the horizons \ref{303}.

Another matter we wished to  investigate was if there was a connection between the reduced symmetry of the multi-horizon solution and its reduced differentiability as alluded to in \cite{Candlish2}  and reviewed here \ref{101}.  If there was such a connection, the three center solution should be even less differentiable than the two center solution. But we have found that in each of the cases  we have studied, \emph{the three center / coplanar solution has an identical  degree  of differentiability as the two center / collinear solution \ref{215}, \ref{224}, \ref{304}}. We have further identified the specific way in which the tensor components are modified when going from the two center situation to the three center situation: (\textbf{P1}) Components which were smooth continue to be  smooth.  (\textbf{P2})  Components which were constant and hence smooth become non-constant with a finite degree of smoothness; but the degree of smoothness is not less than the least degree of smoothness already present in the two center solution.  (\textbf{P3})  Components which had a finite degree of smoothness are modified; but the modifications are such that the degree of smoothness is unchanged.

Now that we have succeeded in settling the three center / coplanar case, the next question would be on the smoothness of the more generic $k$ center solution with $k \geq 4$.  This problem is way more technically involved than the three center problem.  Firstly,  the harmonic function would depend on more isotropic angles viz. $\theta, \phi, \psi_1, \ldots \psi_{k-3}$. We may once again reorganize the harmonic function in terms of appropriate generalized Gegenbauer polynomials as a starting point for the computations: 
\be \label{kcenharm}
H(r,\theta, \phi, \psi_1 \ldots \psi_{k-3}) = \frac{\mu_1}{r^{d-3}} + \sum_{n = 0}^\infty r^n \, {\cal G}_n\,(\theta, \phi, \psi_1, \ldots, \psi_{k-3})\,,
\ee
where ${\cal G}_n\,(\theta, \phi, \psi_1, \ldots, \psi_{k-3})$ are obtained similar to  \eqref{fR} - \eqref{ggpdefn}, but with the $f_i(\theta, \phi)$ replaced by 
the appropriate $f_i(\theta, \phi, \psi_1, \ldots, \psi_{k-3})$.  We would need to solve simultaneously $k$ geodesic equations, the ones for $r(\lambda), \theta(\lambda), \ldots , \psi_{k-3}(\lambda)$. As before if we employ series expansions for these $k$ functions  which means $k$ sets of  infinite coefficients, they would be determined by solving the geodesic equations order by order in a sequence as in \ref{211}.The problem being thus quite involved will not be  considered in its entirety here; we will only make some remarks that follow from the results of this paper.  First let us assume/conjecture that the results of the $k$ center computation are related to the results of the $k-1$ center computation in the same way that the results of the three center computation are related to the results of the two center computation; i.e. the tensor components  follow one of the three possibilities above: \textbf{(P1), (P2), (P3)}. The results of this paper are thus a verification of the above assumption/conjecture for $k=3$. The consequences of this assumption/conjecture for the black hole case (analogous comments hold for $M2$ branes) are: the $d-2$ metric components $g_{v \,\Psi_{k-3}}, g_{\Theta \,\Psi_{k-3}}, g_{\Phi \,\Psi_{k-3}}, g_{\Psi_1 \,\Psi_{k-3}}, \ldots , g_{\Psi_{k-4} \,\Psi_{k-3}}$  and the gauge field component $A_{\Psi_{k-3}}$ follow \textbf{(P2)}, the non-smooth components of the $k-1$ center solution follow \textbf{(P3)} and the rest follow \textbf{(P1)}.   In particular  the degree of smoothness of the $k$ center solution is identical to that of the $k-1$ center solution. 

Now let us take $k$ to be that value for which there are no spatial isometries in the transverse Euclidean space; in the black hole case $k \geq d-1$ and for $M2$ branes $k \geq 8$. Starting with the results of this paper and then successively applying our conjecture, we can arrive at the following statements about the tensor components in the Gaussian null co-ordinate system, for black holes. All components of the metric except the ones restricted by the theory of the co-ordinate system \eqref{281} as well as all components of the gauge field are non-constant,  each with  a degree of differentiability not less than the degree of differentiability of the two center solution. Similar statements hold for the $M2$ brane case. In both cases, \emph{the multi center solution with no spatial isometries has an identical degree of smoothness as the collinear multi center solution}. We aim to perform the computations that will establish our conjecture which leads to the above result on the smoothness for the most generic multi-center solution and report it in a future publication.

 The explicit computations of this paper already spoil the connection with reduced symmetries attributed to the reduced smoothness of multi horizon solutions;  the results that follow from our conjecture, if they are shown to be true, spoil this connection even further. We should note that among the many situations analyzed in \cite{Candlish2} and reviewed here in \ref{101}, we have considered only a subset of them and for this class of examples, there does not seem to be any connection between the reduction of symmetries in the multi horizon solution and the reduction of smoothness. 
 
Instead we offer the following observation;  again pertinent only to the class of examples we have considered here, i.e. $M2$ branes and $d$ dimensional black holes.  \emph{The smoothness of a multi center solution \underline{depends only on the powers of $r$} that occur in the harmonic function that defines the solution}.  From the consideration of powers of $r$ in the harmonic function, the single center solution is different from each $k$ center solution, all of which are identical \eqref{kcenharm}, \eqref{3cenharm3}, \eqref{1.3}. Thus all $k$ center solutions with $k \geq 2$ should have identical degrees of smoothness, different from the degree of smoothness of the single center solution.

 The above statement is a genuine observation only for $k = 1,  2, 3$ both  for   $d \geq 5$  black holes  and for $M2$ branes. The statement itself can be taken as a conjecture (to be verified or disproved) for those cases where computations have not yet been performed viz. $k \geq 4$. 
 
 The above observation/conjecture  involving the powers of $r$ occurring in the harmonic function also covers the $d = 4$ black hole case in the following way. First,  recall that only for $d =4$,  the most general multi center horizon is smooth (thus at the outset, rejecting the connection between symmetries and smoothness.) The $d=4$ case is different from all the $d >4$ cases in the following crucial way. 
The single center is a power ($-1$) of $r$ and the contribution to the harmonic function due to second and higher centers are also powers of $r$ \eqref{kcenharm}.  For $d \neq 4$ however, the single center is a power ($-1$) of not $r$ but of $r^{d-3}$ and the contribution to the harmonic function due to second and higher centers are not powers of $r^{d-3}$ but of $r$ \eqref{kcenharm}. It is in this sense that the powers of $r$ that occur in the harmonic function of $d=4$ is \emph{different} from $d \neq 4$ case;  and our observation/conjecture implies a \emph{difference} in the smoothness between the two.  This difference in power of $r$ in different dimensions for single center is important because it governs the leading order behaviour of $r$ as a function of affine parameter $\lambda$ and hence the parameter for series expansions.  Indeed, it was already shown in \cite{Hartle} that for $d=4$ all horizons are smooth. We can also see it by proceeding along the lines of the present work  (originally \cite{Candlish1}):  we would have ansatz series expansions in $\lambda$ itself and not a fractional power of $\lambda$ as it happens for $d > 4$ and this would then mean that the results of all computations are series in $\lambda$  and hence smooth at $\lambda = 0$. 

Thus our observation/conjecture that \emph{the smoothness of a multi center solution \underline{depends only}  \underline{on the powers of $r$} that occur in the harmonic function that defines the solution}, is meant to hold for all black holes $d \geq 4$ and for $M2$ branes.  We hope to settle the validity or not of this observation/ conjecture with explicit computations in future work. 

Finite differentiability of the metric at the horizon means that an observer falling through the horizon can detect the presence of horizon through  measurement \cite{Chrusciel}. Finite differentiability also  means that some derivatives of the Riemann tensor will blow up at the horizon and these can in principle be observed by an in-falling observer.  We have not studied these mild singularities in a parallel propagating frame in this paper although it is important to do so. The results of this paper and of the conjectures we have made means that the  singularities are in a sense similar for all multi center solutions.  

A natural question which has been asked before is that of the possibility of making these horizons smooth by considering the multi center solutions as solutions to appropriate higher derivative theories.  For the black hole two center solution, this was already attempted in \cite{Candlish2} for a class of higher derivative terms.  From the results of our work and of the conjectures that follow from it,  we can expect that,  if one were to succeed in analyzing the collinear solution, which is considerably easier due to its many spatial isometries, and show that the horizon is smooth in a certain higher derivative theory, the horizons in the coplanar or even the most generic multi center solution will also be smooth. It would also be of interest to study the effect of adding little bit of non-extremality.

For a collinear solution with the centers constituting an infinite periodic array, the horizons turn out to be smooth \cite{Perry}, \cite{Myers}. The results of this paper leads us to speculate that  coplanar solutions constituting infinite periodic arrays on a plane could have smooth horizons. We will leave this interesting question for future work. 

Finally, we are led to the question of the significance, if any, for M-theory physics, of the result of this paper (together with the conjectured consequences that follow from it)  that the metric is  ${\cal C}^3$ and the tensor gauge field ${\cal C}^2$ at the multi $M2$ horizon. For example, via the AdS-CFT correspondence, does it have some implication for appropriate correlators in the dual three dimensional field theories? We will leave these investigations also for the future.

\begin{center}
\textbf{Acknowledgments}
\end{center}

CNG and YKS would like to thank Siddharth Satpathy for initial collaborations. CNG would like to thank the very friendly staff at the various Cafe Coffee Day outlets in Bhubaneshwar, where quite a bit of his contribution to this work was done, for their warm hospitality.

\appendix
\section{\label{a}Solution to geodesic equations in 5d}
Here we provide the solutions to the geodesic equations described in \ref{211}. All the generalized Gegenbauer polynomials ${\cal G}_n$'s appearing in this appendix are the ones relevant for $d = 5$, i.e. constructed out of five dimensional Gegenbauer polynomials. 
\begin{multline}   r(\lambda) = \sqrt{2}\mu_1^{1/4}\,\lambda^{1/2} 			+ \frac{1}{2\sqrt{2}\mu_1^{1/4}}{\cal G}_0 \,\lambda^{3/2} 			+ \frac{2}{5}{\cal G}_1\,\lambda^2  					 -\frac{1}{48 \sqrt{2} \mu _1^{3/4}}\left[ 3\, {\cal G}_0^2 - 32\mu_1\, {\cal G}_2  \right]  \lambda^{5/2} 				-\frac{2}{35 \mu _1^{1/2}}	\left[ {\cal G}_0 \,{\cal G}_1 \right. \\ \left. -10 \mu_1\,{\cal G}_3 \right] \lambda^3  					+\frac{1}{1600 \sqrt{2} \mu _1^{5/4}}  \left[25\, {\cal G}_0^3 - 48 \mu _1\,{\cal G}_1^2  - 2800 \mu _1\, (\partial_\Theta {\cal G}_1)^2 - 2800 \mu _1\, \csc^2\Theta \,(\partial_\Phi {\cal G}_1)^2  \right. \\ \left. +1600\mu _1^2\, {\cal G}_4  \right] \lambda^{\frac72} 				 	+  \frac{1}{126 \mu _1}\left[ 3\,{\cal G}_0^2 G_1  + 12\mu _1\,{\cal G}_0 G_3  -175 \mu _1 \, \partial_\Theta {\cal G}_1 \partial_\Theta {\cal G}_2 -175 \mu _1  \csc^2 \Theta \, \partial_\Phi {\cal G}_1 \partial_\Phi {\cal G}_2 \right. \\ \left. + 112  \mu _1^2\, {\cal G}_5 \right] \lambda^4  				 -\frac{1}{1612800 \sqrt{2} \mu _1^{7/4}} \left[7875\, {\cal G}_0^4  -33600\mu_1\,{\cal G}_0^2  {\cal G}_2  - 42048\mu_1\, {\cal G}_0 {\cal G}_1^2 - 35840\mu _1^2\, {\cal G}_2^2  \right. \\ \left. - 7656768\mu_1\,{\cal G}_0 (\partial_\Theta {\cal G}_1)^2 - 7656768\mu_1\,\csc^2 \Theta\,{\cal G}_0 (\partial_\Phi {\cal G}_1)^2 - 184320\mu _1^2\, {\cal G}_1 {\cal G}_3 - 564480\mu _1^2\, {\cal G}_0 {\cal G}_4  \right. \\ \left.+  967680\mu_1^2\, (\partial_\Theta {\cal G}_2)^2  +  967680\mu_1^2\, \csc^2 \Theta(\partial_\Phi {\cal G}_2)^2  + 2322432\mu_1^2\, \partial_\Theta {\cal G}_1 \partial_\Theta {\cal G}_3 \right. \\ \left. + 2322432\mu_1^2\, \csc^2 \Theta \partial_\Phi {\cal G}_1 \partial_\Phi {\cal G}_3   -  2580480 \mu_1^3\, {\cal G}_6  \right] \lambda^{\frac92}  + \nonumber \end{multline}

\begin{multline}\label{5drsoln} 
  +  \frac{1}{57750 \mu _1^{3/2}} \left[ -600\, {\cal G}_0^3\,+ 294 \mu _1\, {\cal G}_1^3  \right. \\ \left.+ 1500\mu_1 \,{\cal G}_0 \,  {\cal G}_1 \, {\cal G}_2  + 750\mu _1 \,{\cal G}_0^2\, {\cal G}_1  {\cal G}_3 + 279825\mu_1\, {\cal G}_1 \,  (\partial_\Theta {\cal G}_1)^2 + 279825\mu_1\, \csc^2 \Theta\,{\cal G}_1 \,  (\partial_\Phi {\cal G}_1)^2  \right. \\ \left. + 208775\mu_1\, {\cal G}_0\, \partial_\Theta {\cal G}_1\,\partial_\Theta {\cal G}_2   + 208775\mu_1\, \cos^2 \Theta\, {\cal G}_0\, \partial_\Phi {\cal G}_1\,\partial_\Phi {\cal G}_2   - 34125 \mu _1\, ( \partial_\Theta {\cal G}_1)^2 \, \partial^2_\Theta {\cal G}_1 \right. \\ \left.   -  34125 \mu _1\, \csc^4 \Theta\,( \partial_\Phi {\cal G}_1)^2 \, \partial^2_\Phi {\cal G}_1 - 68250 \mu _1\, \csc^2\Theta \,  \partial_\Theta {\cal G}_1\,  \partial_\Phi {\cal G}_1\,\partial^2_{\Theta\Phi} {\cal G}_1 \right. \\ \left. +  34125\mu_1\, \cot \Theta \, \csc ^2 \Theta \, (\partial_\Phi {\cal G}_1)^2 \, \partial_\Theta {\cal G}_1 + 6000\mu_1^2\, {\cal G}_2\, {\cal G}_3 + 12600\mu_1^2\, {\cal G}_1\, {\cal G}_4 + 28000\mu_1^2\, {\cal G}_0\, {\cal G}_5 \right. \\ \left. - 47250 \mu _1^2\, \partial_\Theta {\cal G}_1\, \partial_\Theta {\cal G}_4 - 47250 \mu _1^2\, \csc^2 \Theta\, \partial_\Phi {\cal G}_1\, \partial_\Phi {\cal G}_4  - 38850 \mu _1^2 \, \partial_\Theta {\cal G}_2\, \partial_\Theta {\cal G}_3 \right. \\ \left. - 38850 \mu _1^2 \, \csc^2 \Theta \, \partial_\Phi {\cal G}_2\, \partial_\Phi {\cal G}_3 + 84000  \mu _1^3\,{\cal G}_7 \right] \lambda^5 + \ldots \end{multline}

\begin{multline} \label{5dthetasoln}
\theta(\lambda) = \Theta + \frac{\sqrt{2} }{\mu _1^{1/4}} \partial_\Theta {\cal G}_1\,\lambda^{3/2} + \frac34 \partial_\Theta {\cal G}_2 \,\lambda^2   -\frac{1}{10 \sqrt{2} \mu _1^{3/4}} \left[17 \,{\cal G}_0 \partial_\Theta {\cal G}_1 - 8 \mu _1\, \partial_\Theta {\cal G}_3 \right] \lambda^{5/2} + \frac{1}{20 \mu _1^{1/2}} \left[ -31\, {\cal G}_1 \partial_\Theta {\cal G}_1 \right. \\ \left. + 5\, \partial_\Theta {\cal G}_1 \, \partial^2_\Theta {\cal G}_1 - 10 \, {\cal G}_0\, \partial_\Theta {\cal G}_2 +  15\, \cot \Theta \, \csc ^2 \Theta \, (\partial_\Phi {\cal G}_1)^2 + 5\, \csc ^2 \Theta \, \partial_\Phi {\cal G}_1\, \partial^2_{\Theta\Phi}{\cal G}_1 + 10 \mu _1\, \partial_\Theta {\cal G}_4 \right] \lambda^3  \\  + \frac{1}{560 \sqrt{2} \mu _1^{5/4}} \left[ 937\, {\cal G}_0^2\, \partial_\Theta {\cal G}_1 - 1568\mu _1\, {\cal G}_2 \, \partial_\Theta {\cal G}_1 + 192 \mu _1\, \partial_\Theta {\cal G}_1\,\partial^2_\Theta {\cal G}_2 -576\mu _1\, {\cal G}_1\,  \partial_\Theta {\cal G}_2 - 208\mu_1\, {\cal G}_0 \, \partial_\Theta {\cal G}_3 \right. \\ \left.  + 72 \mu _1\, \partial_\Theta {\cal G}_2\, \partial^2_\Theta {\cal G}_1  + 72 \mu _1\, \csc ^2\Theta\, \partial_\Phi {\cal G}_2 \, \partial^2_{\Theta\Phi} {\cal G}_1 + 576 \mu _1\, \cot \Theta \, \csc ^2\Theta  \, \partial_\Phi {\cal G}_1\, \partial_\Phi {\cal G}_2 \right. \\ \left.  + 192 \mu _1\,\csc ^2 \Theta\,   \partial_\Phi {\cal G}_1 \partial^2_{\Theta\Phi}{\cal G}_2 +  384 \mu _1^2\, \partial_\Theta {\cal G}_5  \right] \lambda^{7/2} + \ldots
\end{multline}
\begin{multline} \label{5dphisoln}
\phi(\lambda) = \Phi + \frac{\sqrt{2}}{\mu _1^{1/4}} \csc^2\Theta\,\partial_\Phi {\cal G}_1\,\lambda^{3/2} + \frac34 \csc^2 \Theta\, \partial_\Phi {\cal G}_2 \,\lambda^2  -\frac{1}{10 \sqrt{2} \mu _1^{3/4}} \csc^2 \Theta \left[17 \,{\cal G}_0 \partial_\Phi {\cal G}_1 - 8 \mu _1\, \partial_\Phi {\cal G}_3 \right] \lambda^{5/2} \\ + \frac{1}{20 \mu _1^{1/2}} \csc ^2 \Theta  \left[ -31\, {\cal G}_1 \, \partial_\Phi {\cal G}_1+  5\, \partial_\Theta {\cal G}_1 \partial^2_{\Theta\Phi}{\cal G}_1 -10\, {\cal G}_0 \,\partial_\Phi {\cal G}_2 + 5 \csc ^2 \Theta \,  \partial_\Phi {\cal G}_1\, \partial^2_\Phi  {\cal G}_1   \right. \\ \left.  - 40 \cot \Theta \, \partial_\Phi {\cal G}_1\, \partial_\Theta {\cal G}_1+ 10 \mu _1\, \partial_\Phi {\cal G}_4\right] \lambda^3  +  \frac{1}{560 \sqrt{2} \mu _1^{5/4}} \csc^2 \Theta \left[  937\, {\cal G}_0^2\,\partial_\Phi {\cal G}_1 - 1568\mu_1\, {\cal G}_2\, \partial_\Phi {\cal G}_1 \right. \\ \left. + 192 \mu _1\, \partial_\Theta {\cal G}_1\, \partial^2_{\Theta\Phi}{\cal G}_2  -  576 \mu_1\, {\cal G}_1 \, \partial_\Phi {\cal G}_2 -208 \mu _1\, {\cal G}_0\,\partial_\Phi {\cal G}_3 +  72 \mu _1\, \partial_\Theta {\cal G}_2\, \partial^2_{\Theta\Phi} {\cal G}_1 \right. \\ \left. + 72 \mu _1\,\csc ^2 \Theta \,  \partial_\Phi {\cal G}_2\, \partial^2_{\Phi}{\cal G}_1 - 960\mu_1\, \cot \Theta \, \partial_\Phi {\cal G}_2\, \partial_\Theta {\cal G}_1 - 720\mu_1\, \cot \Theta \, \partial_\Phi {\cal G}_1\, \partial_\Theta {\cal G}_2 \right. \\ \left. + 192\mu_1\, \csc ^2 \Theta \, \partial_\Phi {\cal G}_1 \, \partial^2_{\Phi} {\cal G}_2  + 384 \mu _1^2 \, \partial_\Phi {\cal G}_5 \right]  \lambda^{7/2} + \ldots \end{multline}
\begin{multline} \label{5dT}
T(\lambda, \Theta, \Phi) = - \frac{\mu_1}{4}\,\lambda^{-1} + \frac{3\mu_1^{1/2}}{4} \, {\cal G}_0 \log \lambda + \frac{8 \sqrt{2}\mu _1^{3/4}}{5}  \,  {\cal G}_1 \lambda^{1/2} + \frac{1}{48} \left[ 33 {\cal G}_0^2 + 80 \mu_1\,{\cal G}_2 \right] \lambda \\ + \frac{2\sqrt{2}\mu _1^{1/4}}{35}\,\left[19\,{\cal G}_0\, {\cal G}_1 + 20\, \mu_1\, {\cal G}_3 \right] \lambda^{3/2} + \frac{1}{400 \mu _1^{1/2}} \left[ 25\, {\cal G}_0^3 + 363 \mu _1\, {\cal G}_1^2 + 750 \mu_1\, {\cal G}_0\,  {\cal G}_2  \right. \\ \left. + 575 \mu _1 \,   (\partial_\Theta {\cal G}_1)^2 + 575 \mu _1 \,\csc^2 \Theta \, (\partial_\Phi {\cal G}_1)^2 + 700\mu _1^2\, {\cal G}_4  \right] \lambda^2 + \frac{1}{630 \sqrt{2} \mu _1^{1/4}} \left[345 \,{\cal G}_0^2\, {\cal G}_1 + 2016\mu _1\, {\cal G}_1 \, {\cal G}_2 \right. \\ \left.  + 2136\mu_1\, {\cal G}_0\, {\cal G}_3  + 1736 \mu _1\, \partial_\Theta {\cal G}_1 \, \partial_\Theta {\cal G}_2  + 1736 \mu _1\, \csc^2 \Theta \, \partial_\Phi {\cal G}_1 \, \partial_\Phi {\cal G}_2  + 1792\mu_1^2\, {\cal G}_5 \right] \lambda^{5/2}  \\ + \frac{1}{100800 \mu _1} \left[-1575 {\cal G}_0^4  + 71400 \mu _1\, {\cal G}_0^2\,{\cal G}_2  + 65124\mu _1 \,{\cal G}_0\, {\cal G}_1^2 + 2772\mu _1\, {\cal G}_0\,  (\partial_\Theta {\cal G}_1)^2 \right. \\ \left.+ 2772\mu _1\, \csc^2\Theta\,{\cal G}_0\,  (\partial_\Phi {\cal G}_1)^2 + 143360\mu_1^2\, {\cal G}_2^2  + 293760\mu_1^2\, {\cal G}_1 \,{\cal G}_3  + 317520\mu _1^2\, {\cal G}_0\, {\cal G}_4  \right. \\ \left. + 60480 \mu _1^2\, (\partial_\Theta {\cal G}_2)^2 +  60480 \mu _1^2\, \csc^2 \Theta (\partial_\Phi {\cal G}_2)^2 + 185472 \mu _1^2\, \partial_\Theta {\cal G}_1\, \partial_\Theta {\cal G}_3 \right. \\ \left. + 185472 \mu _1^2\, \csc^2 \Theta\, \partial_\Phi {\cal G}_1\, \partial_\Phi {\cal G}_3  + 241920 \mu _1^3\, {\cal G}_6 \right] \lambda^3 + \frac{1}{231000 \sqrt{2} \mu _1^{3/4}} \left[ -18525\, {\cal G}_0^3\, {\cal G}_1 + 105264 \mu _1\, {\cal G}_1^3 \right. \\ \left. + 696000\mu_1\, {\cal G}_0 \,{\cal G}_1\,  {\cal G}_2 + 397500 \mu _1 \,{\cal G}_0^2\, {\cal G}_3 - 49200\mu_1\, {\cal G}_1 \, (\partial_\Theta {\cal G}_1)^2 - 49200\mu_1\, \csc^2\Theta\, {\cal G}_1 \, (\partial_\Phi {\cal G}_1)^2 \right. \\ \left. + 204000 \mu _1\, (\partial_\Theta {\cal G}_1)^2\, \partial^2_\Theta {\cal G}_1 + 204000 \mu _1\, \csc^4\Theta \,(\partial_\Phi {\cal G}_1)^2\, \partial^2_\Phi {\cal G}_1 + 241000 \mu_1\, {\cal G}_0\,  \partial_\Theta {\cal G}_1\, \partial_\Theta {\cal G}_2 \right. \\ \left. + 241000 \mu_1\,\csc^2 \Theta \, {\cal G}_0\,  \partial_\Phi {\cal G}_1\, \partial_\Phi {\cal G}_2 + 408000\mu_1\,  \csc ^2 \Theta \,  \partial_\Phi {\cal G}_1\, \partial_\Theta {\cal G}_1\,  \partial^2_{\Theta \Phi}{\cal G}_1\right. \\ \left.  - 204000 \mu_1 \cot \Theta \, \csc ^2 \Theta \,  (\partial_\Phi {\cal G}_1)^2\,\partial_\Theta  {\cal G}_1 + 1200000\mu _1^2\, {\cal G}_2\, {\cal G}_3  + 1252800\mu _1^2 \, {\cal G}_1\, {\cal G}_4  + 1376000\mu _1^2\, {\cal G}_0 \, {\cal G}_5  \right. \\ \left. +  648000 \mu _1^2 \, \partial_\Theta {\cal G}_1\, \partial_\Theta {\cal G}_4 +  648000 \mu _1^2 \,\csc^2\Theta \,\partial_\Phi {\cal G}_1\, \partial_\Phi {\cal G}_4 + 348000 \mu _1^2\, \partial_\Theta {\cal G}_2\,  \partial_\Theta {\cal G}_3 \right. \\ \left. + 348000 \mu _1^2\, \csc^2 \Theta \, \partial_\Phi {\cal G}_2\,  \partial_\Phi {\cal G}_3  + 960000\mu _1^3\, {\cal G}_7 \right] \lambda^{7/2} + \ldots  \end{multline}

\section{\label{b}Solution to geodesic equations in $d \geq 6$}
Here we provide the solutions to the geodesic equations described in \ref{221}. All the generalized Gegenbauer polynomials ${\cal G}_n$'s appearing in this appendix are the ones relevant for generic $d > 5$, i.e. constructed out of $d > 5$ dimensional Gegenbauer polynomials. 
\begin{multline} \label{6drsoln} r(\lambda) = (d-3)^{1/{d-3}}\, \mu_1^{\frac{d-4}{(d-3)^2}}\, \lambda^{1/{d-3}} + \sum_{l=-3}^{d-7} \frac{d-4}{2d+l-3}\,(d-3)^{\frac{l+4}{d-3}}\, \mu_1^{\frac{d(l+3)-4l-13}{(d-3)^2}}\,{\cal G}_{l+3} \, \left( \lambda^{1/{d-3}}  \right)^{d+l+1} \\ + \frac{d-4}{8\, (d-3)^3 \, (2 d+k-3)} \, (d-3)^{\frac{2 d-5}{d-3}}\,\mu _1^{-\frac{d-2}{(d-3)^2}}\,\left[(d-2) (d-6) \,{\cal G}_0^2 + 8 (d-3)^2 \mu _1 \,{\cal G}_{k+3} \right] (\lambda^{1/{d-3}})^{2 d - 5} \\ + \frac{d-4}{2 (d-3)^3 (2 d-5) (2 d+k-2)} (d-3)^{\frac{2 d-4}{d-3}}\mu _1^{-\frac{2}{(d-3)^2}} \left[ (d-1) \left(d^2-8 d+14\right) {\cal G}_0\, {\cal G}_1 \right. \\ \left. + 2 (d-3)^2 (2 d-5) \mu_1\,{\cal G}_{k+4} \right] (\lambda^{1/{d-3}})^{2d - 4} + \ldots \end{multline}
\begin{multline} \label{6dthetasoln} \theta(\lambda)  = \Theta + \sum_{l = 0}^{d-4} \frac{d-2}{(d+l-2)\,(l+1)}\,(d-3)^{\frac{l+1}{d-3}}\,\mu_1^{\frac{(d-4)l-1}{(d-3)^2}}\,\partial_\Theta {\cal G}_{l+1}\,(\lambda^{1/{d-3}})^{l+d-2} + \ldots \end{multline}
\begin{multline} \label{6dphisoln} \phi(\lambda)  = \Phi + \sum_{l = 0}^{d-4} \frac{d-2}{(d+l-2)\,(l+1)}\,(d-3)^{\frac{l+1}{d-3}}\,\mu_1^{\frac{(d-4)l-1}{(d-3)^2}}\,\csc^2 \Theta \, \partial_\Phi {\cal G}_{l+1}\,(\lambda^{1/{d-3}})^{l+d-2} + \ldots \end{multline}
\begin{multline} \label{6dT}  T(\lambda, \Theta, \Phi) =   - \frac{1}{(d-3)^2}\,\mu _1^{2/{d-3}}\, \lambda^{-1} \left[1  -     (d -2)\,\mu _1^{-1/{d-3}}\,{\cal G}_0 \, \lambda \,\log \lambda \right. \\  \left. - \sum_{l=1}^{d-4}   \frac{2 d + 2 l - 4}{l\,(2 d + l - 6)}\,(d-3)^{2+ \frac{l}{d-3}} \,\mu_1^{\frac{l(d-4)-d+3}{(d-3)^2}}\,{\cal G}_l \, \lambda^{\frac{d + l -3}{d-3}} + \ldots \right]  \end{multline}

\section{\label{c}Solution to geodesic equations in 11d}
Here we provide the solutions to the geodesic equations described in \ref{301}. All the generalized Gegenbauer polynomials ${\cal G}_n$'s appearing in this appendix are the ones relevant for the eleven dimensional membrane harmonic function, i.e. constructed out of $d = 9$ dimensional Gegenbauer polynomials. 

\begin{multline}  r(\lambda) = \sqrt{2}\mu_1^{1/{12}}\lambda^{1/2} + \frac{1}{2\sqrt{2}\mu_1^{5/{12}}}{\cal G}_0\,\lambda^{7/2} + \frac{8}{27 \mu _1^{1/3}}\,{\cal G}_1 \,\lambda^4 +\frac{4\sqrt{2} }{15 \mu _1^{1/4}} {\cal G}_2\,  \lambda^{9/2}  + \frac{16}{33 \mu _1^{1/6}} {\cal G}_3\, \lambda^5\\ + \frac{4 \sqrt{2}}{9 \mu _1^{1/{12}}} {\cal G}_4\, \lambda^{{11}/2} + \frac{32}{39} {\cal G}_5\, \lambda^6   + \frac{1}{252 \sqrt{2} \mu _1^{11/12}} \left[ - 119 \, {\cal G}_0^2 + 384 \mu _1 {\cal G}_6 \right] \lambda^{{13}/2} + \frac{32}{405 \mu _1^{5/6}} \left[ -11 {\cal G}_0 \,{\cal G}_1 + 18\mu_1\, {\cal G}_7 \right] \lambda^7 \\ + \frac{\sqrt{2}}{127575 \mu _1^{3/4}} \left[ -102060\, {\cal G}_0 \,{\cal G}_2 - 51625\, {\cal G}_1^2 - 2673 \, (\partial_\Theta {\cal G}_1)^2 - 2673 \, \csc^2 \Theta \, (\partial_\Phi {\cal G}_1)^2 + 170100\mu _1\, {\cal G}_8 \right] \lambda^{{15}/2} \\ +\frac{4}{58905 \mu _1^{2/3}}   \left[- 21700\,  {\cal G}_0\,  {\cal G}_3 - 22176\,  {\cal G}_1 \,{\cal G}_2 - 935\, \partial_\Theta {\cal G}_1\, \partial_\Theta {\cal G}_2 - 935 \csc^2 \Theta \, \partial_\Phi {\cal G}_1\, \partial_\Phi {\cal G}_2 + 36960 \mu _1 \,{\cal G}_9 \right] \lambda^8 \\ + \frac{2 \sqrt{2}}{467775 \mu _1^{7/12}}\left[ -165396 \,{\cal G}_2^2 - 327600\, {\cal G}_1\, {\cal G}_3- 317625 {\cal G}_0\, {\cal G}_4 - 5775 (\partial_\Theta {\cal G}_2)^2 - 11440 \, \partial_\Theta {\cal G}_1\, \partial_\Theta {\cal G}_3 \right. \\ \left.  - 5775 \csc^2 \Theta (\partial_\Phi {\cal G}_2)^2 - 11440 \csc^2\Theta \, \partial_\Phi {\cal G}_1\, \partial_\Phi {\cal G}_3 + 554400 \mu _1\, {\cal G}_{10}\right] \lambda^{{17}/2}  + \frac{8}{23108085 \mu _1^{1/2}}\left[ - 7665840\,  {\cal G}_2\, {\cal G}_3 \right. \\ \left.  - 7527520 \, {\cal G}_1\, {\cal G}_4 - 7234920 \, {\cal G}_0\,  {\cal G}_5  - 225225\, \partial_\Theta {\cal G}_2\, \partial_\Theta {\cal G}_3 - 220077 \, \partial_\Theta {\cal G}_1\, \partial_\Theta {\cal G}_4 - 225225\, \csc^2 \Theta \, \partial_\Phi {\cal G}_2\, \partial_\Phi {\cal G}_3 \right. \\ \left. - 220077 \,\csc^2 \Theta\, \partial_\Phi {\cal G}_1\, \partial_\Phi {\cal G}_4  + 12972960 \mu _1 \,{\cal G}_{11} \right] \lambda^9  + \nonumber \end{multline} 

\begin{multline} \label{11drsoln}+\frac{1}{1248647400 \sqrt{2} \mu _1^{17/12}}\left[ 1534988455 \,{\cal G}_0^3 \right. \\ \left.  - 3133428480 \mu _1\,  {\cal G}_3^2  - 6215489280 \mu _1\,  {\cal G}_2\, {\cal G}_4 - 6056117760 \mu _1\, {\cal G}_1\, {\cal G}_5  - 5771525760 \mu _1\, {\cal G}_0\, {\cal G}_6   \right. \\ \left.  - 78524160 \mu _1\,  (\partial_\Theta {\cal G}_3)^2 -  155387232 \mu _1 \, \partial_\Theta {\cal G}_2\, \partial_\Theta {\cal G}_4 - 149140992 \mu _1\, \partial_\Theta {\cal G}_1\,  \partial_\Theta {\cal G}_5  - 78524160 \mu _1\,  \csc^2 \Theta \, (\partial_\Phi {\cal G}_3)^2 \right. \\ \left.  -  155387232 \mu _1 \,  \csc^2 \Theta \,\partial_\Phi {\cal G}_2\, \partial_\Phi {\cal G}_4 - 149140992 \mu _1\,  \csc^2 \Theta \, \partial_\Phi {\cal G}_1\,  \partial_\Phi {\cal G}_5  + 10655124480 \mu _1^2\, {\cal G}_{12}\right] \lambda^{{19}/2} \\ +  \frac{16}{42567525 \mu _1^{4/3}} \left[ 9214205\, {\cal G}_0^2 \, {\cal G}_1 - 12612600 \mu _1\,  {\cal G}_3\,  {\cal G}_4 - 12418560 \mu _1\,  {\cal G}_2\,  {\cal G}_5 - 12012000 \mu _1\,  {\cal G}_1\,   {\cal G}_6  \right. \\ \left.- 11351340  \mu _1 \, {\cal G}_0\,{\cal G}_7 - 272415 \mu _1\,  \partial_\Theta {\cal G}_3\, \partial_\Theta {\cal G}_4 + 266175 \mu _1 \,\partial_\Theta {\cal G}_2\, \partial_\Theta {\cal G}_5 - 249678   \mu _1\, \partial_\Theta {\cal G}_1\, \partial_\Theta  {\cal G}_6 \right. \\ \left.  - 272415 \mu _1\, \csc^2\Theta \, \partial_\Phi {\cal G}_3\, \partial_\Phi {\cal G}_4 + 266175 \mu _1 \, \csc^2 \Theta \, \partial_\Phi {\cal G}_2\, \partial_\Phi {\cal G}_5 - 249678 \mu _1\, \csc^2 \Theta \, \partial_\Phi {\cal G}_1\, \partial_\Phi  {\cal G}_6 \right. \\ \left.  + 21621600 \mu _1^2\, {\cal G}_{13} \right] \lambda^{10} + \frac{1}{120405285 \sqrt{2} \mu _1^{5/4}} \left[ 781539759\,{\cal G}_0^2\, {\cal G}_2 + 788124337 \, {\cal G}_0 \, {\cal G}_1^2  + 29930553 {\cal G}_0\,  ( \partial_\Theta {\cal G}_1)^2   \right. \\ \left.  + 29930553 {\cal G}_0\,  \csc^2 \Theta \, ( \partial_\Phi {\cal G}_1)^2 - 542702160 \mu _1\, {\cal G}_4^2  - 1077753600 \mu _1 \, {\cal G}_3\, {\cal G}_5 - 1054145664 \mu _1\,  {\cal G}_2 \,  {\cal G}_6  \right. \\ \left. - 1012467456 \mu _1\,  {\cal G}_1\, {\cal G}_7  - 948647700  \mu _1\, {\cal G}_0 \, {\cal G}_8  - 10216206 \mu _1\, (\partial_\Theta {\cal G}_4)^2 - 20217600 \mu _1\,  \partial_\Theta {\cal G}_3\,  \partial_\Theta {\cal G}_5 \right. \\ \left. - 19459440 \mu _1\,  \partial_\Theta {\cal G}_2\, \partial_\Theta {\cal G}_6 - 17729280 \mu _1\,  \partial_\Theta {\cal G}_1\, \partial_\Theta {\cal G}_7 - 10216206 \mu _1\, \csc^2 \Theta \, (\partial_\Phi {\cal G}_4)^2 \right. \\ \left. - 20217600 \mu _1\, \csc^2 \Theta \, \partial_\Phi {\cal G}_3\,  \partial_\Phi {\cal G}_5  - 19459440 \mu _1\, \csc^2 \Theta \,  \partial_\Phi {\cal G}_2\, \partial_\Phi {\cal G}_6 - 17729280 \mu _1\, \csc^2 \Theta \,  \partial_\Theta {\cal G}_1\, \partial_\Theta {\cal G}_7 \right. \\ \left. + 1868106240 \mu _1^2 \,{\cal G}_{14} \right] \lambda^{{21}/2} + \ldots \end{multline}

\begin{multline}  \theta(\lambda) = \Theta + \frac{4 \sqrt{2}}{35 \mu _1^{5/12}}\partial_\Theta {\cal G}_1  \, \lambda^{7/2} + \frac{1}{6 \mu _1^{1/3}} \partial_\Theta {\cal G}_2\,\lambda^4 + \frac{8 \sqrt{2} }{63 \mu _1^{1/4}} \partial_\Theta {\cal G}_3\,\lambda^{9/2}   +\frac{1}{5 \mu _1^{1/6}} \partial_\Theta {\cal G}_4\,  \lambda^5 + \frac{16 \sqrt{2}}{99 \mu_1^{1/{12}}}   \partial_\Theta {\cal G}_5\, \lambda^{{11}/2} \\ + \frac{4}{15} \partial_\Theta {\cal G}_6\,  \lambda^6  + \frac{2 \sqrt{2}}{2145 \mu _1^{11/12}} \left[  - 241\, {\cal G}_0\,\partial_\Theta {\cal G}_1 + 240 \mu _1 \,\partial_\Theta {\cal G}_7\right] \lambda^{{13}/2} + \frac{2}{33075 \mu _1^{5/6}}  \left[-6790 \, {\cal G}_1\,  \partial_\Theta {\cal G}_1 \right. \\ \left. - 5775\,  {\cal G}_0\,  \partial_\Theta {\cal G}_2  + 90 \,\partial_\Theta {\cal G}_1 \, \partial^2_\Theta {\cal G}_1 +  90\,  \csc ^2\Theta \, \partial_\Phi {\cal G}_1 \, \partial^2_{\Theta\Phi}{\cal G}_1 + 126 \cot \Theta \, \csc ^2 \Theta \,  (\partial_\Phi {\cal G}_1)^2 + 6300 \mu _1 \, \partial_\Theta {\cal G}_8 \right]\lambda^7 \\ +\frac{2 \sqrt{2}}{184275 \mu _1^{3/4}} \left[ - 34776 \, {\cal G}_2 \, \partial_\Theta {\cal G}_1 - 29680\, {\cal G}_1\, \partial_\Theta {\cal G}_2 - 25650\, {\cal G}_0\,  \partial_\Theta {\cal G}_3 + 315 \, \partial_\Theta {\cal G}_2\, \partial^2_\Theta {\cal G}_1 + 432 \, \partial_\Theta {\cal G}_1 \, \partial^2_\Theta {\cal G}_2 \right. \\ \left. + 315\,  \csc ^2 \Theta \, \partial_ \Phi {\cal G}_2\,  \partial^2_{\Theta \Phi} {\cal G}_1 + 432 \csc^2 \Theta \, \partial_\Phi {\cal G}_1\,  \partial^2_{\Theta\Phi}{\cal G}_2 + 1008\,  \cot \Theta \,  \csc ^2 \Theta  \,  \partial_\Phi {\cal G}_1\, \partial_\Phi {\cal G}_2 + 30240 \mu _1\, \partial_\Theta {\cal G}_9\right] \lambda^{{15}/2} \\ + \frac{1}{582120 \mu _1^{2/3}}\left[  - 406224 \, {\cal G}_3\, \partial_\Theta {\cal G}_1 - 347424\,  {\cal G}_2\,  \partial_\Theta  {\cal G}_2 - 301840 \, {\cal G}_1\, 
   \partial_\Theta {\cal G}_3 - 263340 \, {\cal G}_0\, \partial_\Theta {\cal G}_4 \right. \\ \left. + 2640 \, \partial_\Theta {\cal G}_3\, \partial^2_\Theta {\cal G}_1 +  3465 \, \partial_\Theta {\cal G}_2\,  \partial^2_\Theta {\cal G}_2 + 4752\, \partial_\Theta  {\cal G}_1 \, \partial^2_\Theta {\cal G}_3 + 2640 \, \csc^2 \Theta \, \partial_\Phi {\cal G}_3\, \partial^2_{\Theta\Phi} {\cal G}_1 \right. \\ \left. + 3465\, \csc^2 \Theta \, \partial_{\Phi}{\cal G}_2\,  \partial^2_{\Theta\Phi}{\cal G}_2 + 4752\, \csc^2 \Theta \, \partial_\Phi {\cal G}_1\,  \partial^2_{\Theta\Phi} {\cal G}_3 + 4620\, \cot \Theta \, \csc^2 \Theta\, (\partial_\Phi {\cal G}_2)^2  \right. \\ \left. +  9504\,  \cot  \Theta \csc^2 \Theta \,  \partial_\Phi {\cal G}_1\, \partial_\Phi {\cal G}_3 + 332640 \mu _1 \, \partial_\Theta {\cal G}_{10} \right] \lambda^8 + \nonumber \end{multline}
   
  \begin{multline} \label{11dthetasoln} + \frac{4 \sqrt{2}}{883575 \mu _1^{7/12}}\left[ -143220 \,{\cal G}_4\,\partial_\Theta {\cal G}_1 - 122640\,  {\cal G}_3 \partial_\Theta {\cal G}_2 \right. \\ \left.  - 106920 \, {\cal G}_2\,   \partial_\Theta {\cal G}_3 - 93940\, {\cal G}_1\, \partial_\Theta {\cal G}_4 - 82390 \, {\cal G}_0\,  \partial_\Theta {\cal G}_5 + 693 \, \partial_\Theta {\cal G}_4 \,  \partial^2_\Theta {\cal G}_1 + 880\, \partial_\Theta {\cal G}_3\,  \partial^2_\Theta {\cal G}_2 \right. \\ \left. + 1155\,  \partial_\Theta {\cal G}_2\,  \partial^2_\Theta {\cal G}_3 + 1584 \, \partial_\Theta {\cal G}_1\,  \partial^2_\Theta {\cal G}_4 + 693\, \csc^2 \Theta \, \partial_\Phi {\cal G}_4\, \partial^2_{\Theta\Phi}{\cal G}_1  + 880\, \csc^2 \Theta \,  \partial_\Phi {\cal G}_3\,  \partial^2_{\Theta\Phi} {\cal G}_2  \right. \\ \left.  +  1155\, \csc^2 \Theta \,  \partial_\Phi {\cal G}_2\,  \partial^2_{\Theta\Phi} {\cal G}_3 + 1584\, \csc^2 \Theta \,  \partial_\Phi {\cal G}_1\, \partial^2_{\Theta\Phi} {\cal G}_4 +  2640\,  \cot \Theta \, \csc^2 \Theta \,  \partial_\Phi {\cal G}_2\,  \partial_\Phi {\cal G}_3 \right. \\ \left.  + 2772 \,  \cot \Theta \, \csc^2 \Theta \,  \partial_\Phi {\cal G}_1\, \partial_\Phi  {\cal G}_4 + 110880 \mu _1 \, \partial_\Theta {\cal G}_{11} \right] \lambda^{{17}/2} + \ldots \end{multline}

\begin{multline}  \label{11dphisoln}   \phi(\lambda) = \Phi + \frac{4 \sqrt{2}}{35 \mu _1^{5/12}}\csc^2\Theta \,\partial_\Phi {\cal G}_1\, \lambda^{7/2} + \frac{1}{6 \mu _1^{1/3}} \csc^2\Theta \,\partial_\Phi {\cal G}_2\, \lambda^4 + \frac{8 \sqrt{2} }{63 \mu _1^{1/4}}\csc^2\Theta \, \partial_\Phi {\cal G}_3\, \lambda^{9/2} \\   +\frac{1}{5 \mu _1^{1/6}} \csc^2\Theta \, \partial_\Phi {\cal G}_4\,   \lambda^5 + \frac{16 \sqrt{2}}{99 \mu_1^{1/{12}}} \csc^2\Theta \,  \partial_\Phi {\cal G}_5\, \lambda^{{11}/2} + \frac{4}{15} \csc^2\Theta \,\partial_\Phi {\cal G}_6\, \lambda^6  + \frac{2 \sqrt{2}}{2145 \mu _1^{11/12}}\csc^2\Theta \, \left[  - 241\, {\cal G}_0 \,\partial_\Phi {\cal G}_1 \right. \\ \left.  +  240 \mu _1 \,\partial_\Phi {\cal G}_7 \right] \lambda^{{13}/2} + \frac{2}{33075 \mu _1^{5/6}} \csc^2 \Theta\,\left[- 6790\,  {\cal G}_1\, \partial_\Phi {\cal G}_1 - 5775\,  {\cal G}_0 \, \partial_\Phi {\cal G}_2 + 90 \, \partial_\Theta {\cal G}_1\,  \partial^2_{\Theta \Phi} {\cal G}_1 \right. \\ \left. +  90\, \csc ^2\Theta\,   \partial_\Phi {\cal G}_1\, \partial^2_\Phi {\cal G}_1 - 432 \,  \cot  \Theta  \,  \partial_\Phi {\cal G}_1\,  \partial_\Theta {\cal G}_1 + 6300 \mu _1 \, \partial_\Phi {\cal G}_8 \right] \lambda^7 + \frac{2 \sqrt{2}}{184275 \mu _1^{3/4}} \csc^2 \Theta\,  \left[ - 34776\,  {\cal G}_2\,\partial_\Phi {\cal G}_1 \right. \\ \left.  - 29680 \, {\cal G}_1\,\partial_\Phi {\cal G}_2 - 25650\,  {\cal G}_0 \, \partial_\Phi {\cal G}_3 + 315\, \partial_\Theta {\cal G}_2\, \partial^2_{\Theta \Phi}{\cal G}_1 +  432 \, \partial_\Theta {\cal G}_1 \, \partial^2_{\Theta \Phi} {\cal G}_2  + 315 \, \csc^2 \Theta \, \partial_\Phi {\cal G}_2\, \partial^2_\Phi {\cal G}_1 \right. \\ \left. +  432\, \csc^2 \Theta \,  \partial_\Phi {\cal G}_1\,  \partial^2_{\Phi}{\cal G}_2  - 1872 \, \cot  \Theta \,  \partial_\Phi {\cal G}_2\,  \partial_\Theta {\cal G}_1 - 1638 \, \cot \Theta \, \partial_\Phi {\cal G}_1 \, \partial_\Theta  {\cal G}_2  + 30240 \mu _1 \, \partial_\Phi {\cal G}_9 \right] \lambda^{{15}/2}   \\ +  \frac{1}{582120 \mu _1^{2/3}}\,\csc^2\Theta \left[ -406224\, {\cal G}_3\, \partial_\Phi {\cal G}_1 - 347424\,  {\cal G}_2 \,  \partial_\Phi {\cal G}_2 - 301840 \, {\cal G}_1 \, 
   \partial_\Phi {\cal G}_3 - 263340 \, {\cal G}_0 \, \partial_\Phi {\cal G}_4 \right. \\ \left. + 2640\,  \partial_\Theta {\cal G}_3\,  \partial^2_{\Theta\Phi}{\cal G}_1 + 3465 \, \partial_\Theta {\cal G}_2\, \partial^2_{\Theta\Phi}{\cal G}_2 + 4752\,  \partial_\Theta {\cal G}_1\, \partial^2_{\Theta\Phi} {\cal G}_3 +  2640 \, \csc^2 \Theta \, \partial_\Phi {\cal G}_3 \, \partial^2_{\Phi} {\cal G}_1 \right. \\ \left. +  3465\,  \csc^2 \Theta \, \partial_\Phi {\cal G}_2 \, \partial^2_{\Phi} {\cal G}_2 + 4752 \, \csc^2 \Theta\, \partial_\Phi {\cal G}_1\,  \partial^2_{\Phi} {\cal G}_3  - 19008 \, \cot  \Theta \,  \partial_\Phi {\cal G}_3 \,  \partial_\Theta {\cal G}_1 -  16170 \, \cot \Theta \,  \partial_\Phi {\cal G}_2 \,  \partial_\Theta {\cal G}_2 \right. \\ \left.  - 14784\, \cot \Theta \,  \partial_\Phi {\cal G}_1\,  \partial_\Theta {\cal G}_3 + 332640 \mu _1\, \partial_\Phi {\cal G}_{10} \right] \lambda^8 + \frac{4 \sqrt{2}}{883575 \mu _1^{7/12}} \csc^2 \Theta\,\left[  - 143220\, {\cal G}_4\, \partial_\Phi {\cal G}_1\right. \\ \left.  - 122640 \, {\cal G}_3\,  \partial_\Phi {\cal G}_2 - 106920 \, {\cal G}_2 \,    \partial_\Phi {\cal G}_3 - 93940\,  {\cal G}_1 \,  \partial_\Phi {\cal G}_4 - 82390\,  {\cal G}_0 \, \partial_\Phi {\cal G}_5 + 693 \, \partial_\Theta {\cal G}_4\,  \partial^2_{\Theta \Phi}{\cal G}_1 \right. \\ \left. + 880\,  \partial_\Theta {\cal G}_3\, \partial^2_{\Theta \Phi}{\cal G}_2 + 1155\, \partial_\Theta {\cal G}_2 \,  \partial^2_{\Theta \Phi}{\cal G}_3 + 1584 \, \partial_\Theta {\cal G}_1\,  \partial^2_{\Theta \Phi}{\cal G}_4 + 693 \,\csc^2 \Theta\, \partial_\Phi {\cal G}_4\,  \partial^2_{ \Phi}{\cal G}_1 \right. \\ \left.  + 880\, \csc^2 \Theta\, \partial_\Phi {\cal G}_3\, \partial^2_{\Phi}{\cal G}_2 + 1155\,\csc^2 \Theta\, \partial_\Phi {\cal G}_2 \,  \partial^2_{ \Phi}{\cal G}_3 + 1584 \, \csc^2 \Theta\,\partial_\Phi {\cal G}_1\,  \partial^2_{\Phi}{\cal G}_4  - 5940 \, \cot \Theta \, \partial_\Phi {\cal G}_4\, \partial_\Theta {\cal G}_1 \right. \\ \left. - 4950\,  \cot \Theta \, \partial_\Phi {\cal G}_3 \,  \partial_\Theta {\cal G}_2 - 4400 \, \cot  \Theta \,  \partial_\Phi {\cal G}_2 \, \partial_\Theta {\cal G}_3 - 4158\,  \cot  \Theta \,  \partial_\Phi {\cal G}_1\,  \partial_\Theta {\cal G}_4 + 110880 \mu _1 \, \partial_\Phi {\cal G}_{11} \right] \lambda^{{17}/2} + \ldots  \end{multline}

\begin{multline} \label{11dT} T(\lambda, \Theta, \Phi) = -\frac{\mu _1^{1/3}}{4}\lambda^{-1} + \frac{7 }{12 \mu _1^{1/6}} {\cal G}_0\,\lambda^2 + \frac{64 \sqrt{2} }{135 \mu _1^{1/{12}}} {\cal G}_1 \lambda^{5/2} + \frac{4}{5} {\cal G}_2\,\lambda^3 + \frac{160 \sqrt{2} \mu _1^{1/{12}}}{231}  {\cal G}_3 \, \lambda^{7/2} + \frac{11 \mu _1^{1/6}}{9}  {\cal G}_4\,\lambda^4 \\ + \frac{128\sqrt{2} \mu _1^{1/4}}{117}  {\cal G}_5 \, \lambda^{9/2} + \frac{1}{1260 \mu _1^{2/3}} \left[ -259 \,{\cal G}_0^2 + 2496 \mu _1\,  {\cal G}_6 \right] \lambda^5 + \frac{224 \sqrt{2}}{4455 \mu _1^{7/12}}\left[-7\, {\cal G}_0\, {\cal G}_1 + 36 \mu _1\, {\cal G}_7 \right] \lambda^{{11}/2}\\ + \frac{1}{255150 \mu _1^{1/2}}\left[ -78925 \,{\cal G}_1^2 - 153090\, {\cal G}_0\, {\cal G}_2 + 13851 \, (\partial_\Theta {\cal G}_1)^2 + 13851 \, \csc^2 \Theta \, (\partial_\Phi {\cal G}_1)^2 + 850500 \mu _1\, {\cal G}_8 \right] \lambda^6  \\ + \frac{16 \sqrt{2}}{765765 \mu _1^{5/12}} \left[- 25872\,  {\cal G}_1\, {\cal G}_2 - 24325 \, {\cal G}_0 \, {\cal G}_3 + 4114\,  \partial_\Theta {\cal G}_1\, \partial_\Theta {\cal G}_2 + 4114\, \csc^2 \Theta\,   \partial_\Phi {\cal G}_1\, \partial_\Phi {\cal G}_2 \right. \\ \left.  + 147840 \mu _1\, {\cal G}_9 \right]  \lambda^{{13}/2} + \frac{2}{3274425 \mu _1^{1/3}}\left[ -790944\, {\cal G}_2^2 - 1537200 \, {\cal G}_1\, {\cal G}_3 - 1397550\, {\cal G}_0 \,{\cal G}_4 + 109725 \,(\partial_\Theta {\cal G}_2)^2 \right. \\ \left. + 233200\,  \partial_\Theta {\cal G}_1\, \partial_\Theta {\cal G}_3 + 109725 \,\csc^2 \Theta \, (\partial_\Phi {\cal G}_2)^2  + 233200\,\csc^2 \Theta \,  \partial_\Phi {\cal G}_1\, \partial_\Phi {\cal G}_3 + 9424800 \mu _1 \, {\cal G}_{10}\right] \lambda^7 \\ + \frac{16 \sqrt{2}}{115540425 \mu _1^{1/4}} \left[-6191640\, {\cal G}_2\, {\cal G}_3 - 5845840 \, {\cal G}_1 \,  {\cal G}_4 - 5114340 \, {\cal G}_0 \, {\cal G}_5 + 791505\,  \partial_\Theta {\cal G}_2 \, \partial_\Theta {\cal G}_3 \right. \\ \left. + 880308\,  \partial_\Theta {\cal G}_1 \, \partial_\Theta {\cal G}_4  + 791505\, \csc^2\Theta\, \partial_\Phi {\cal G}_2 \, \partial_\Phi {\cal G}_3  + 880308\, \csc^2 \Theta \,  \partial_\Phi {\cal G}_1 \, \partial_\Phi {\cal G}_4 + 38918880 \mu _1 \, {\cal G}_{11} \right] \lambda^{{15}/2} \\ + \frac{1}{624323700 \mu _1^{7/6}}\left[ 196931735\, {\cal G}_0^3 - 484553160 \mu _1\, {\cal G}_3^2 - 943422480 \mu _1\,  {\cal G}_2 \,  {\cal G}_4 - 863736720 \mu _1\,  {\cal G}_1\, {\cal G}_5 \right. \\ \left. - 721440720 \mu _1 \, {\cal G}_0\,  {\cal G}_6 + 55306680 \mu _1 \, (\partial_\Theta {\cal G}_3)^2 + 115846731 \mu _1\,  \partial_\Theta {\cal G}_2\,  \partial_\Theta {\cal G}_4 + 133429296 \mu _1\, 
   \partial_\Theta {\cal G}_1\,  \partial_\Theta {\cal G}_5 \right. \\ \left. + 55306680 \mu _1 \, \csc^2 \Theta \,  (\partial_\Phi {\cal G}_3)^2 + 115846731 \mu _1\,  \partial_\Phi {\cal G}_2\,  \partial_\Phi  {\cal G}_4 + 133429296 \mu _1\,   \partial_\Phi {\cal G}_1\,  \partial_\Phi {\cal G}_5 \right. \\ \left.  + 6326480160 \mu _1^2 \,{\cal G}_{12} \right] \lambda^8 
  +  \frac{8 \sqrt{2}}{723647925 \mu _1^{13/12}}\left[77322245\, {\cal G}_0^2 \,{\cal G}_1 - 126126000 \mu _1\, {\cal G}_3\, {\cal G}_4 \right. \\ \left. - 119528640 \mu _1\,  {\cal G}_2 \,  {\cal G}_5 - 105705600 \mu _1\,  {\cal G}_1\, {\cal G}_6 - 83243160 \mu_1\,{\cal G}_0\, {\cal G}_7  + 13427700 \mu _1\,  \partial_\Theta {\cal G}_3\, \partial_\Theta {\cal G}_4 \right. \\ \left. + 14578200 \mu _1 \, \partial_\Theta {\cal G}_2 \, \partial_\Theta {\cal G}_5 + 17256096 \mu _1\,  \partial_\Theta {\cal G}_1\,  \partial_\Theta {\cal G}_6 + 13427700 \mu _1\, \csc^2\Theta \,  \partial_\Phi {\cal G}_3\, \partial_\Phi {\cal G}_4 \right. \\ \left.  + 14578200 \mu _1 \, \csc^2 \Theta \, \partial_\Phi {\cal G}_2 \, \partial_\Phi {\cal G}_5 + 17256096 \mu _1\, \csc^2 \Theta \,  \partial_\Phi {\cal G}_1\,  \partial_\Phi {\cal G}_6  + 864864000 \mu _1^2\,  {\cal G}_{13} \right] \lambda^{{17}/2} + \ldots \end{multline}

\section{\label{d}Regularity condition}

We begin with \eqref{fghconstraint}, the condition that the $f, g, h$ need to satisfy to have the metric to be non-singular on the horizon:
\be \label{detcond}   f^2 (q_6^2 - z_2 z_3) + g^2 (q_4^2 - z_3 z_1) + h^2 (q_2^2 - z_1 z_2)   + 2 f g (q_4 q_6 - q_2 z_3) - 2 g h (q_2 q_4 - q_6 z_1) +  2 f h (q_6 q_2 - q_4 z_2)  \neq 0. \ee 
This is the determinant of the following $4\times 4$ matrix (the metric for the ``AdS'' part), evaluated at the horizon i.e. at $\lambda=0$.
\begin{equation}
M = \left (\begin{array}{cccc}
0 & -g & -h & f \\
-g & z_2 & q_6 & q_2 \\
-h & q_6 & z_3 & q_4 \\
f & q_2 & q_4 & z_1 \\ \end{array} \right)
\end{equation}
The determinant of $M$, given by \ref{detcond} can be written as a quadratic form
\be \det(M) = -F^{T}S F \ee with 
\be F= \left(\begin{array}{ccc}
-g & -h & f  \end{array} \right), \qquad S  =
\left (\begin{array}{ccc}
 A & D & G \\
 D & E & H \\
 G & H & I \\ \end{array} \right).
\end{equation}
Here matrix $S$ is the adjugate (or classical adjoint) corresponding to lower $3\times 3$ block in $M$. It's a singular matrix. We can see why this matrix appears as follows. If $g_{ab}$ is a symmetric matrix then following identity holds
\begin{equation}
\det \left (\begin{array}{ccc}
x & v_b\\
w_a & g_{ab} \\ \end{array} \right) = (x -v_b g^{bc}w_{c}) \det[g_{ab}].
\end{equation}
Applying this identity to matrix $M$, we get that $\det[M] = -F^{T}S F$. Notice that lower $3\times 3$ block in $M$ i.e. $g_{ab}$ is a singular matrix
 whose inverse $g^{ab}$ is not defined. 
\begin{equation}
g= \left (\begin{array}{ccc}
 z_2 & q_6 & q_2 \\
q_6 & z_3 & q_4 \\
 q_2 & q_4 & z_1 \\ \end{array} \right) = \left (\begin{array}{ccc}
 f_X & g_X & h_X \\
f_Y & g_Y & h_Y \\
 f_v& g_v & h_v \\ \end{array} \right)\left (\begin{array}{ccc}
 -f_X & -f_Y & -f_v \\
g_X & g_Y & g_v \\
 h_X & h_Y & h_v \\ \end{array} \right)
\end{equation}

Here subscript $X,Y,v$ below $f,g,h$ denotes partial derivative with respect to that variable. Since there is a functional relation between $f,g$ and $h$, the determinant of these matrices vanish individually. To put it another way, take various partial derivatives of relation $f^2 -g^2 -h^2 =1$ and  demand a non-trivial solution for the resulting linear equations. But $g^{ab}Det[g_{ab}]$ still gives the adjugate matrix $S$. We can think of first working with a non-zero $\lambda$ so that inverse is well defined and then taking the limit $\lambda \rightarrow 0$. 

We will analyze this matrix $S$ in detail. Matrix elements are as follows:
\ben
A&=& z_1z_3 -q_{4}^{2}, \quad D= q_2q_4 - z_1 q_6, \quad E=  z_1z_2 -q_{2}^{2} \nonumber \\
I&=& z_2z_3 -q_{6}^{2}, \quad H= q_2q_6 - z_2 q_4, \quad G= q_6q_4 - z_3q_2 \
\een
Elements of $S$ satisfy some identities which can be easily checked:
\ben \label{identities}
AE -D^2 = z_1 \det[g] ,  \quad IA- G^2 = z_3  \det[g],  \quad DG-AH= q_4 \det[g]  \nonumber \\  EI-H^2 = z_2 \det[g],  \quad DH-GE= q_2 \det[g] ,  \quad GH-ID= q_6 \det[g] \nonumber \\ Az_2 + Dq_6 + Gq_2 = \det[g],  \quad Dz_2 + Eq_6 + Hq_2 =0 ,  \quad Gz_2 + Hq_6 + Iq_2 =0
\een
In our case $\det[g]=0$ and so some of the above identities simplify. Eigenvalues $\alpha$ of $S$ are given by the characteristic equation
\be
\det[S] -c_2 \alpha + Tr(S)\alpha^2 -\alpha^3 =0
\ee
where Cayley-Hamilton theorem gives 
\be
c_2 = \frac{1}{2}\left( (Tr(S))^2 - Tr(S^2)\right) = AE + AI + EI -(D^2+G^2+H^2).
\ee
Since $\det[S]=0$ and by the above identities, $c_2=0$, we have only one non-zero eigenvalue $\alpha= Tr(S)$.  Define $3$-vectors 
\be
\vec{K} = \nabla g \times \nabla h,\quad \vec{L} = \nabla f \times \nabla g,\quad \vec{M} = \nabla f \times \nabla h 
\ee
In terms of these, we can write, after little bit of algebra,
\be
z_2z_3 -q_6^2 = K_v^2 -L_v^2 -M_v^2,\quad z_1z_3 -q_4^2 = K_X^2 -L_X^2 -M_X^2,\quad z_2z_1 -q_2^2 = K_Y^2 -L_Y^2 -M_Y^2.
\ee
So the non-zero eigenvalue 
\be
\alpha = Tr(S) =A+ E+ I = \vec{K}\cdot \vec{K} -\vec{L}\cdot \vec{L}-\vec{M}\cdot \vec{M}
\ee
Using the condition \ref{detcond}, we can write $\vec{L} = \frac{-h}{f}\vec{K}$ and $\vec{M} = \frac{g}{f}\vec{K}$ since $\nabla f = \frac{g}{f}\nabla g + \frac{h}{f}\nabla h$. Using this, we see that $\alpha = \frac{\vec{K}\cdot\vec{K}}{f^2}$. So if $\vec{K}$ is zero then all eigenvalues are zero and hence the quadratic form vanishes. We can determine the eigenvector corresponding to $\alpha$. This is given as a solution to 
\be
\left (\begin{array}{ccc}
 A & D & G \\
 D & E & H \\
 G & H & I \\ \end{array} \right)\left(\begin{array}{ccc}
x_1 & x_2 & x_3
                  \end{array} \right) = \alpha \left(\begin{array}{ccc}
x_1 & x_2 & x_3
                  \end{array} \right)
\ee
Since $A,E,I$ have same sign, we can write $D=\sqrt{AE}$, $G=\sqrt{AI}$ and $H=\sqrt{EI}$. Then eigenvector is $(\sqrt{A},\sqrt{E},\sqrt{I}$. If we use the matrix of eigenvectors to diagonalize $S$ then quadratic form can be written as 
\be
\det[M]= (f\sqrt{A} -g\sqrt{E} - h\sqrt{I})^2 =  
(f\sqrt{(-q_6^2 + z_2 z_3)} - g\sqrt{(-q_4^2 + z_1 z_3)} -h\sqrt{(-q_2^2 + z_2 z_1)})^2. 
\ee
Now we know that a quadratic form which doesn't change sign (since $\det[M]$ is always positive, this condition is satisfied for our case) can only be zero on the null space of the matrix $S$. So we can have determinant $M$ zero only if $F$ belongs to null space of $S$. We check that unless $f,g,h$ are zero, this is not the case. An example of eigenvector with zero eigenvalue, using identities \ref{identities} is $(z_2,q_6,q_2)$.

So the only way for determinant of $M$ to vanish is if $\vec{K} = \nabla g \times \nabla h =0 $. We can see that if $f,g,h$ are constants or if all are functions of only one variable then $\det[M]$ would vanish and metric would be singular. Some possible choices of $f,g,h$ are given in \eqref{fghexamples}.

\end{document}